\documentclass{jfm}

\usepackage{graphicx}
\usepackage{natbib}
\usepackage{amsmath}
\usepackage{amssymb}
\usepackage{multirow}
\usepackage{xcolor}
\usepackage{nomencl}

\newcommand{\av}[1]{\overline{#1}}
\newcommand{\fl}[1]{{#1}'}
\renewcommand{\d}[0]{\textnormal{d}}
\newcommand{\Ri}[0]{\textit{Ri}}
\renewcommand{\Re}[0]{\textit{Re}}

\newcommand{\ignore}[1]{}

\newcommand{\rev}[1]{{#1}}

\title[Mixing and entrainment are suppressed
in inclined gravity currents]{Mixing and entrainment are suppressed \\
in inclined gravity currents}

\author[M. van Reeuwijk et al.]{Maarten van Reeuwijk$^1$, Markus
  Holzner$^2$, C. P.  Caulfield$^{3,4}$}

\affiliation{$^1$Department of Civil and Environmental Engineering, Imperial College London, London SW7 2AZ, UK
\\[\affilskip]
$^2$Institute of Environmental Engineering, ETH Z\"{u}rich, CH-8039 Z\"{u}rich, Switzerland\\[\affilskip]
$^3$BP Institute, University of Cambridge, Madingley Rise, Madingley
Road, Cambridge CB3 0EZ, UK \\
$^4$Department of Applied Mathematics \& Theoretical Physics,
University of Cambridge, Wilberforce
Road, Cambridge CB3 0WA, UK}

\begin{document}

\maketitle

\begin{abstract}
We explore the dynamics of inclined temporal gravity currents using direct numerical simulation, and find that the current creates an environment in which the flux Richardson number $\Ri_f$, gradient Richardson number $\Ri_g$, and turbulent flux coefficient $\Gamma$ are constant across a large portion of the depth.
Changing the slope angle $\alpha$ modifies these mixing parameters, and the flow approaches a maximum Richardson number $\Ri_\textrm{max}\approx 0.15$ as $\alpha \rightarrow 0$ at which the entrainment coefficient $E \rightarrow 0$.
The turbulent Prandtl number remains $O(1)$ for all slope angles, demonstrating that $E\rightarrow 0$ is not caused by a switch-off of the turbulent buoyancy flux as conjectured by Ellison (1957). 
Instead, $E \rightarrow 0$ occurs as the result of the turbulence intensity going to zero as $\alpha\rightarrow 0$, due to the flow requiring larger and larger shear to maintain the same level of turbulence.
We develop an approximate model valid for small $\alpha$ which is able to predict accurately $\Ri_f$, $\Ri_g$ and $\Gamma$ as a function of $\alpha$ and their maximum attainable values. 
The model predicts an entrainment law of the form $E=0.31(\Ri_\textrm{max}-\Ri)$, which is in good agreement with the simulation data.
The simulations and model presented here contribute to a growing body of evidence that an approach to a marginally \rev{or critically} stable, relatively weakly stratified equilibrium for  stratified shear flows may well be  a generic property of turbulent stratified flows. 

\end{abstract}

\section{Introduction}
\label{sec:intro}
Gravity currents are a regular occurrence in nature, e.g.\ katabatic
winds, dense downslope releases in the ocean, pyroclastic flows and
ventilation exchange flows between spaces of differing temperatures \citep{Simpson1999}.
As it propagates, a  gravity current exerts a shear on the ambient
fluid which,
for flows at sufficiently
large Reynolds number,  consequently leads to turbulence production
and mixing, causing ambient fluid to be entrained into the current, generically increasing its characteristic depth $h$.
Simultaneously, since the current is more dense than the ambient fluid, such stratification typically suppresses mixing.
Therefore, the appropriate nondimensional measure of the rate of entrainment, i.e. the entrainment parameter $E$ (defined in detail below),
should be a function of an appropriate bulk
`Richardson number'
$\Ri$ (once again defined in detail below) quantifying
the relative importance of the potential
energy increase associated
with the deepening of the current
compared to the kinetic energy of the propagating current.

Turbulent entrainment in inclined gravity currents was first studied experimentally  by \cite{Ellison1959}.
Their experimental geometry comprised an inclined channel in which a fluid lighter (heavier) than the ambient was injected which flowed along the channel top (bottom) as a gravity current.
By varying the channel inclination angle and thus the bulk Richardson
number $\Ri$, an entrainment law of the form $E = f(Ri)$ was observed,
which was subsequently parameterised \citep{Turner1986} by
\begin{equation}
E=\frac{(0.08-0.1 \Ri)}{(1+5 \Ri)}. \label{eq:et59}
\end{equation}

Crucially, this parameterisation has at its heart that there is a
critical Richardson number at which $E \rightarrow 0$, and
so assumes that sufficiently strong stratification `switches off' the
entrainment.
However, field campaigns of oceanic overflows and several new
experimental investigations have since revealed orders of magnitudes difference in the observed values of $E$ \citep[e.g.][]{Wells2010},
highlighting a need for further understanding of the physical
processes responsible for turbulent entrainment.
As discussed by \cite{Wells2010}, the process of shear-driven
entrainment associated with gravity-current-like outflows is one of
the key processes determining diapycnal transport in the world's
oceans \citep{Ferrari2009}.

\rev{It} is becoming increasingly apparent
that mixing in the vicinity of basin boundaries plays an essential role
in the meridional overturning circulation of the world's oceans
\citep{Ferrari2016}.
\rev{Entrainment and mixing associated with gravity currents play a key role
  in such boundary-located mixing and the associated development
  of stratification within ocean basins. Evidence from laboratory
  experiments
  and simplified bulk models  \citep[see for example][]{Wells2005,Wahlin2006} show that entrainment and mixing in gravity
  currents on slopes can both be
  strongly sensitive to the slope angle, and also can (naturally) have
  a leading order effect on the large-scale basin
  stratification. As such currents can have a significant horizontal
  extent, and in at least some circumstances can be relatively long-lived,} 
it  is timely to consider in controlled
circumstances the shear-driven entrainment processes at the top
of a propagating gravity current.
\rev{Crucially, such a current is free of the organised `head' dynamics \citep[see][for more details]{Sher2015}, which dominate the  entrainment for starting gravity currents and lock exchange problems.}
  
A particular area of remaining controversy, dating
from the seminal work of \cite{Ellison1959},  is
 whether sufficiently strong stratification
does indeed
`switch off' turbulent entrainment completely.
Comparison between various studies is made
more challenging due to the wide range
of possible definitions of Richardson number.
It is also important to remember that the processes involved in
stratified shear-driven 
turbulent mixing are inherently three-dimensional, \rev{(not least because of the central role played by  secondary  instabilities in triggering a forward 
cascade to smaller scales in the vorticity field)} so the
computation
of such flows is highly resource-intensive.  
However, there is some recent numerical evidence that suggests that, at least in some highly
controlled and idealised circumstances, stratified
turbulence cannot be sustained at sufficiently high Richardson number.

\cite{Deusebio2015} considered stratified plane Couette flow, i.e.\ the flow in a plane channel of depth $2H$ between two
horizontal planes which are maintained at constant
(statically stable and different) density $\rho_a \mp \rho_0$ and velocity $\pm U_0$. (Here,
$\rho_a$ is a reference density and the so-called Boussinesq
approximation is valid, in that $\rho_0 \ll \rho_a$.)
As
well as the Prandtl number   $Pr=\nu/\kappa$ ($\nu$ being the kinematic
viscosity and $\kappa$ being the diffusivity of the density
respectively)   this flow
has two key parameters: the
bulk Richardson number $Ri_0$,
and the Reynolds number $Re$, here defined
as
\begin{equation}
Ri_0= \frac{g \rho_0 H}{\rho_a U_0^2 }, \
Re= \frac{U_0 H}{\nu } .\label{eq:spcparam}
\end{equation}

Such flows have the particular attraction that
they satisfy the underlying  assumptions of Monin-Obukhov (M-O) similarity
theory \citep{Obukhov1971,Monin1954}, namely
that the boundary conditions enforce
 a constant (vertical) flux of
buoyancy throughout the channel gap.
Effectively, there is a constant-with-depth vertical transport of
density determined by the imposed boundary conditions.  \cite{Deusebio2015}
observed that, consistent with this theory, the equilibrium streamwise
flow velocity $u_e$ and density
distribution $\rho_e$ (corresponding
to horizontal and temporal averages in the simulations)
approached self-similar (and inherently coupled) functional forms
well-described by the M-O theory. Importantly, these coupled
profiles naturally lead to an `equilibrium' Richardson
number
\begin{equation}
Ri_e= \frac{-\frac{g}{\rho_a} \frac{d \rho_e}{dz}}{\left ( \frac{d
      u_e}{dz}
\right )^2} , \label{eq:rie}
\end{equation}
which is approximately constant, at least
 sufficiently
far from the boundaries, an observation that is robust for a range
of flow Prandtl numbers, as demonstrated
by \cite{Zhou2017}. Significantly, (see for example figure 18 of
\cite{Deusebio2015} and figure 7a of \cite{Zhou2017})  $Ri_e \lesssim 0.2$ for arbitrarily large external Reynolds
number. Dynamically, no matter how high the external \rev{i.e.\ associated with the wall forcing)} Reynolds number
is, turbulence (and associated
turbulent mixing and vertical transport) cannot be sustained
for large stratification.
The turbulence becomes intermittent,
thus
suggesting that, at least in such a self-similar and idealised flow,
turbulence (and the ensuing vertical transport) can be `switched off'
by strong stratification.

Further evidence for this `switch-off' is provided in \cite{Krug2017}, who study the fractal properties of the turbulent-nonturbulent interface (TNTI) using data from Direct Numerical Simulation (DNS) of a temporal inclined gravity current \citep{vanReeuwijk2017}. In this study, it was found that the fractal dimension $\beta$ of the TNTI depends linearly on the gradient Richardson number $\Ri_b$, as $\beta = 1.60 (\Ri_g - 0.15)$, thus suggesting a critical $\Ri_g$ at 0.15 (note that this is the 2D fractal dimension, i.e.\ based on analysis of the line contour of two-dimensional transects).
This observation was linked to the anisotropy in the horizontal and
vertical length scales, with horizontal lengths scaling with the layer
thickness $h$ and the vertical lengths scaling  with the shear length scale $e^{1/2}/S$, where $e$ is the turbulence kinetic energy and $S$ is the strain rate (again defined below).
Critically, the anisotropy in the length scales was \emph{not} related
to anisotropy in the turbulence itself -- thus clearly suggesting that
the `switch-off' is not related to a transition to very strongly
stratified or `layered anisotropic stratified turbulence' regime \citep{Falder2016}
discussed in \cite{Brethouwer2007} and based on the scaling and
theoretical
arguments of \cite{Billant2001} and \cite{Lindborg2006}.

Although this `switch-off' in itself is apparently consistent with the
pioneering modelling work of \cite{Ellison1957}, it is important to
appreciate that the conceptual picture is qualitatively different. Ellison's vision, reasonably based upon the data available at the time,
was that `turbulence can be maintained at large values of $\Ri$',
and so, for internal consistency when considering
mixing processes, it was necessary that the turbulent Prandtl number
$Pr_T=\nu_T/\kappa_T \rightarrow \infty$ in such very stable
conditions,
so that the mixing and entrainment effects of that turbulence could
be `switched off' even when the turbulence is itself active.
Not only in the stratified plane Couette flow discussed above, but
also in homogeneous sheared stratified turbulence \citep[e.g. see][and
references therein]{Shih2005} and inflectional shear layers unstable
primarily either to Kelvin-Helmholtz instability \citep{Salehipour2015}
 or especially to Holmboe wave instability \citep{Salehipour2018}, there is increasing evidence that $Pr_T$ does not diverge, but rather
asymptotes to an $O(1)$ quantity, and the ultimate
suppression of mixing by stratification arises because
of the (as yet empirical)
observation that
 there is an equilibrium or stationary Richardson number
beyond
which turbulence cannot be sustained.
\rev{These qualitatively different conceptual pictures of how strong
  stratification might `switch off' entrainment and mixing make it
  necessary to define carefully what actually is meant by such
  a `switch off', as discussed further below.}

Of course, not least because of the fact that the observed critical value of this equilibrium Richardson number is close to the critical Richardson number ($1/4$) of the Miles-Howard criterion \citep{Miles1961,Howard1961} for the linear stability of steady laminar parallel inviscid stratified shear flow, it has been conjectured that stratified flows adjust to a state
of `marginal stability' with Richardson number close to this critical
value \citep{Thorpe2009}.
There is observational evidence in the Equatorial Undercurrent \citep{Smyth2013} that the distribution of measured Richardson numbers
is peaked around $1/4$, which is certainly consistent with the
`marginal stability' conjecture, although it is
also consistent with the picture that turbulence switches off beyond a
certain critical value. 
In particular, for shear flows with sufficiently `sharp' and strong
initial 
stratification so that the flow is primarily susceptible to the
Holmboe wave instability, there is strong evidence 
\citep[see][for further details]{Salehipour2018}  that the ensuing
turbulence exhibits `self-organized criticality' \citep{Bak1987}, such that
the PDF of a notional Richardson number defined using horizontal
averages of streamwise velocity and density becomes strongly peaked
around $1/4$, irrespective of the initial conditions.
Whether this critical value is set by linear
instability processes is, it is fair to note, at the moment unclear.
Just to mention two open issues, in stratified plane Couette flow, the
turbulence, when it becomes intermittent cannot really
be characterised by an onset of a linear instability process,
while it is also not yet explained why stability calculations based around
profiles from
horizontal averages of strongly turbulent flows might be well-posed
and relevant \rev{to the ensuing statistically steady flow dynamics}.

The central objective of this paper is to explore in detail the
underlying physical reasons for the observed apparent `switch-off' in entrainment.
To address this objective, we revisit the temporal gravity current
simulations of \cite{vanReeuwijk2017, Krug2017}, and organise the rest
of the paper as follows. In section \ref{sec:setup} we describe the 
set-up of our three-dimensional numerical simulations. In section \ref{sec:ss}, 
we demonstrate that the flows exhibit self-similarity, and define
various appropriate measures of irreversible mixing and entrainment for these flows. 
In section \ref{sec:turb}, we then develop a turbulence
parameterisation, focussing in particular on the turbulent Prandtl
number and  
an approximate model for small angle. Finally, in sections
\ref{sec:disc} and \ref{sec:conc} we discuss our results, and draw
brief conclusions.
\rev{A nomenclature is provided at the end of the paper}.

\section{Case setup}\label{sec:setup}

\begin{figure}
\centering
\includegraphics[width=10cm]{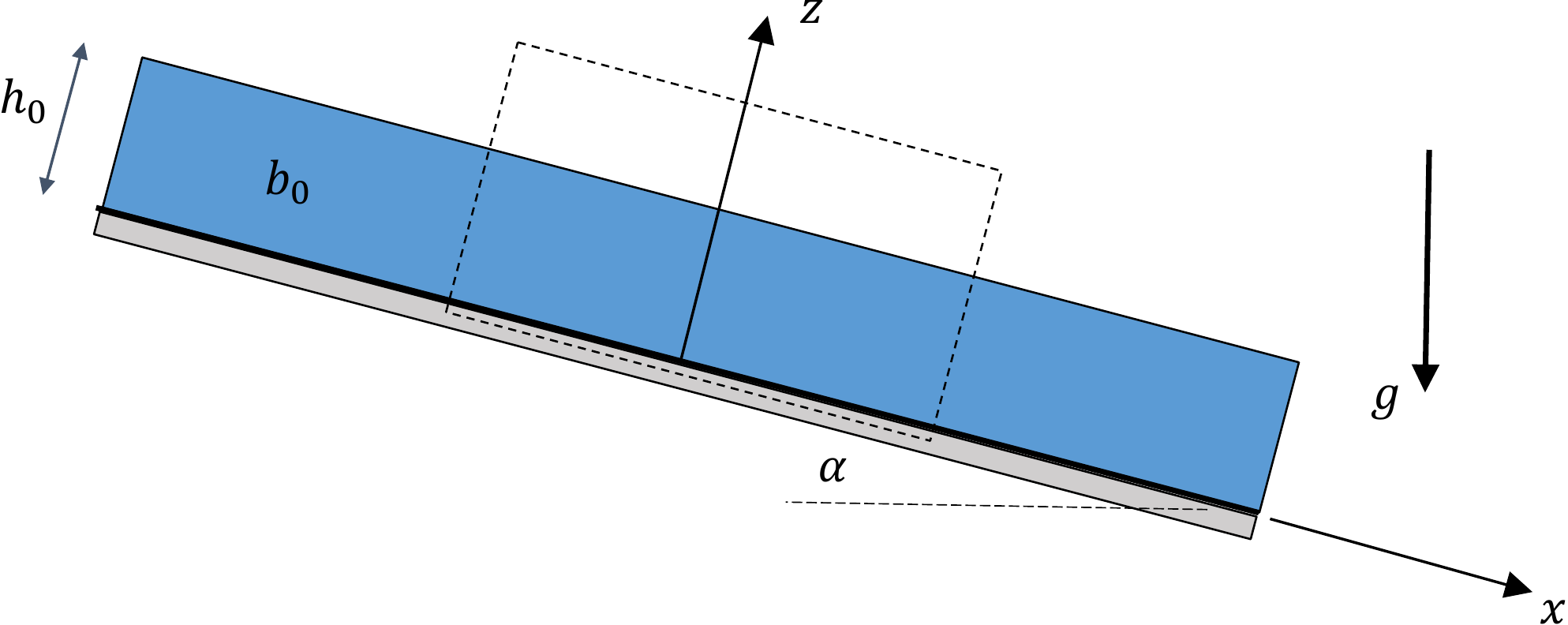}
\caption{Schematic representation of the simulation setup.}
\label{fig:sketch}
\end{figure}

The simulations comprise a temporal version of the classical inclined gravity current experiments of \citet{Ellison1959, Krug2013}, as outlined in \cite{vanReeuwijk2017}.
Specifically, the problem entails a negatively buoyant (heavy) fluid
layer of infinite extent flowing down a slope of angle $\alpha$, as
shown schematically in figure \ref{fig:sketch}.
The dense fluid layer has an initial buoyancy $b_0<0$ and velocity $U_0$.
Here, buoyancy is defined as $b = g (\rho_a - \rho) / \rho_a$ where
$g$ is the gravitational acceleration and $\rho_a$ is the ambient
density of the quiescent layer above the current.
For an angle $\alpha = 90^o$, this case represents a plane wall plume.
\rev{In order to simulate this unbounded problem, a finite size domain is selected (dotted lines in Figure \ref{fig:sketch}) and periodic boundary conditions are imposed on the lateral boundaries in the streamwise and spanwise directions.}
Because of the flow geometry and the periodic boundary conditions, the flow will remain statistically
homogeneous in the $x$ and $y$ direction, and its statistics thus only
depend on the wall-normal coordinate $z$ and time $t$ \citep[see
also][]{Fedorovich2009}. As noted in \S \ref{sec:intro}, this geometry
removes any `head' dynamics of the gravity current, and focusses
exclusively on the shear-induced entrainment of ambient fluid at the
top of the current.
At the bottom wall, a no-flux (Neumann) boundary condition is enforced for buoyancy.
For velocity, both no-slip and free-slip conditions will be considered.

\rev{The difference between a temporal and a spatial gravity current is that in the former the problem is spatially homogeneous and evolves in time, whilst in the latter, the problem is spatially inhomogeneous but steady in time.
This has implications for the entrainment behaviour and the average position of the turbulent-nonturbulent interface (TNTI). In the temporal case, the TNTI's average position evolves in time and the mean velocity through the TNTI is zero. In the spatial case, the TNTI's average position is fixed and there is a mean velocity through the TNTI.
However, these two problems are dynamically very similar as the physics of turbulent entrainment is governed by the \emph{difference} between the TNTI velocity and the flow velocity \citep{DaSilva2014}.}

The characteristic velocity scale, layer thickness and buoyancy, $u_T$, $h$, and $b_T$, respectively, are defined as
\begin{equation}
 \label{eq:integrals2}
 u_T h  = \int_0^\infty \av{u} \d z, \quad \quad
 u_T^2 h = \int_0^\infty \av{u}^2 \d z, \quad \quad
 b_T h =\int_0^\infty \av{b} \d z,
\end{equation}
where the overline denotes averaging over the (homogeneous) $x$ and $y$ directions.
We note that the integral buoyancy forcing $B_0$, defined
as
\begin{equation}
  B_0=  - b_T h \sin \alpha = - b_0 h_0 \sin \alpha,
  \label{eq:b0def}
\end{equation}
is a \rev{positive} conserved quantity for this flow, and that $h$,  $u_T$ and $b_T$ are expected to scale as $h \sim B_0^{1/2} t $, $u_T \sim B_0^{1/2}$, $b_T \sim B_0^{1/2} t^{-1}$  \citep{vanReeuwijk2017}.
The flow cases were designed such that $B_0$ was identical for all angles; this ensures that the integral forcing in the streamwise momentum equation is identical for all flow cases considered, thus clearly bringing out any differences in the turbulence structure.
\rev{Here, we are interested in flows where the turbulence (and
  associated mixing and entrainment) are driven purely by buoyancy
  effects, and so we are interested in how the flow dynamics vary as
  $\alpha$
  varies, and in particular as it approaches small, yet crucially
  always finite, values. Since we  have chosen to keep $B_0$ constant
  across
  all cases, our formulation does not allow for the consideration
  of $\alpha=0$ precisely. However, we aim to identify scalings as
  $\alpha$ takes very small values, in particular to understand the
  properties of the turbulence, entrainment and mixing in that
  limit. It is also important to remember at such shallow angles other
  physical processes, such as large-scale pressure
gradients, hydraulic controls, or indeed substantial bottom roughness,
are likely to be much more significant in determining the flow
dynamics in geophysically-relevant situations. Here, we are focussed on understanding the properties of
buoyancy-driven shear-induced turbulence in such temporal gravity
currents.}

The simulations are carried out with the direct numerical simulation
code SPARKLE, which solves the Navier-Stokes equations in the Boussinesq approximation and is fully parallelised making use of domain decomposition in two directions.
The spatial differential operators are discretised using second order symmetry-preserving central differences \citep{Verstappen2003}, and time-integration is carried out with an adaptive second order Adams-Bashforth method \citep{vanReeuwijk2008a}.
Periodic boundary conditions are applied for the lateral directions.

The (bulk) Reynolds number $\Re$ and bulk Richardson number $\Ri$ are defined as
\begin{equation}
  \Re = \frac{u_T h_0}{\nu}, \quad \quad \Ri =  - \frac{b_T h \cos
    \alpha}{u_T^2},
\label{eq:reridef}
\end{equation}
Consistent with \citep{vanReeuwijk2017, Krug2017}, the simulations were performed at $\Pr=1$ and initial Reynolds number $\Re_0 = U_0 h_0 /\nu = 3890$  on a large domain of $20 h_0 \times 20 h_0 \times 10 h_0$ to ensure reliable statistics for this transient problem.
A resolution of $N_x \times N_y \times N_z = 1536^2 \times 1152$,
sufficient for (fully three-dimensional) direct numerical simulation, is employed for all simulations.

\renewcommand{\arraystretch}{1.5}
\begin{table}
\caption{\label{tab:simdata}Simulation data. Simulation domain  for all simulations is $20 h_0 \times 20 h_0 \times 10 h_0$ at a resolution of $1536^2 \times 1152$. NS, FS: no-slip and free-slip velocity boundary conditions, respectively. }
\centering
\begin{tabular}{l|cccccccccc}
Sim.\ & $\alpha$ & BC & $\Re_0$ & $\Ri_0$ & $t_{run}/t^{\star}$ & $t_{stat}/t^{\star}$ & $\Re_b$ & $
\Re_\lambda$ & $\Delta x / \eta_K$ \\
\hline
  S2 &        2 & FS &     3890 &     0.50 &       55 &       10 &      308 &       75 &     1.47 \\
  S5 &        5 & FS &     3890 &     0.20 &       40 &       12 &      687 &       99 &     1.41 \\
 S10 &       10 & FS &     3890 &     0.10 &       40 &       12 &     1276 &      131 &     1.16 \\
S10N &       10 & NS &     3815 &     0.10 &       40 &       12 &     1019 &       78 &     1.18 \\
 S25 &       25 & FS &     3890 &     0.04 &       25 &        4 &     2132 &      145 &     1.06 \\
 S45 &       45 & FS &     3890 &     0.02 &       20 &        6 &     5437 &      152 &     1.09 \\
 S90 &       90 & FS &     3890 &     0.00 &       20 &        6 & $\infty$ &      166 &     1.08 \\
\end{tabular}
\renewcommand{\arraystretch}{1.0}
\end{table}

Further simulation details can be found in table \ref{tab:simdata}.
The simulations were performed for a duration $t_{run}$ and statistics were calculated over an interval $t_{stat}$ at the end of the simulation.
Interestingly, over this period the flow is fully self-similar, and
the profiles of $\overline{u}$ and $\overline{b}$ are very close to linear in $z$ \citep{vanReeuwijk2017}, consistent with laboratory experiments \citep{Odier2009, Krug2015, Odier2014}.
The typical time scale is defined as $t^*=h_0 / \sqrt{B_0}$.
The grid is chosen such that $\Delta x / \eta_K < 1.5$ for all flow cases.
Here $\eta_K=(\nu^3/\varepsilon_T)^{1/4}$ is the Kolmogorov lengthscale where $\varepsilon_T \equiv h^{-1} \int \varepsilon \d z$ is the characteristic dissipation rate.
Also shown in table \ref{tab:simdata} are the buoyancy Reynolds number $\Re_b$ and Taylor Reynolds number $\Re_\lambda$, defined as
\begin{equation}
\Re_b = \frac{\varepsilon_T}{\nu \hat{N}^2}, \quad \quad \Re_\lambda = \frac{u'_T \lambda_T}{\nu}.
\end{equation}
Here $N^2 = \d \overline{b} / \d z \cos \alpha$  is the square buoyancy frequency
where $(\hat{\cdot})$ denotes averaging over the `outer layer' interval $z/h \in [1/2, 1]$, $\lambda_T = \sqrt{10 \nu e_T / \varepsilon_T}$ is the Taylor length scale and $u'_T=\sqrt{2 e_T/3}$ where $e_T \equiv h^{-1} \int e \d z$ is the characteristic turbulence kinetic energy \citep[][pp. 67-68]{Tennekes1972}.
Both $\Re_b$ and $\Re_\lambda$ become constant as a function of time, and the values presented here are the typical values once a fully self-similar flow evolution is established.
\rev{For an impression of the flow field and a characterisation of the turbulent-nonturbulent interface, see \cite{Krug2017}}.

\section{Self-similarity and measures of mixing}\label{sec:ss}

\subsection{Dependence on boundary conditions}
Before considering all the cases, we first focus on the mixing behaviour for  $\alpha=10^o$ and examine the differences between no-slip boundary conditions (simulation S10N) and free-slip boundary conditions (S10).
In particular, we will demonstrate: 1) that the outer layer behaviours of S10 and S10N are practically indistinguishable with respect to their mixing behaviour; and 2) that the mixing characteristics are practically uniform in the outer layer of the flow -- something which seems to be a unique feature of temporal gravity currents.

It is clear that free-slip boundaries do not occur in nature, but if the
entrainment
between the current and the ambient fluid is dominated by outer
(effectively shear-driven) processes, the influence of velocity
boundary conditions should be small. We do not address here the important
question of the interplay between
turbulence generated by bottom roughness and turbulence
generated by shear-driven processes
at the interface between the
current and ambient (see \cite{Wells2010} for further discussion),
\rev{or indeed driving mechanisms of a current other than the buoyancy
  force associated with a finite (albeit potentially exceptionally
  small) slope angle.}

\begin{figure}
\centering
\hspace{-3mm}
\includegraphics{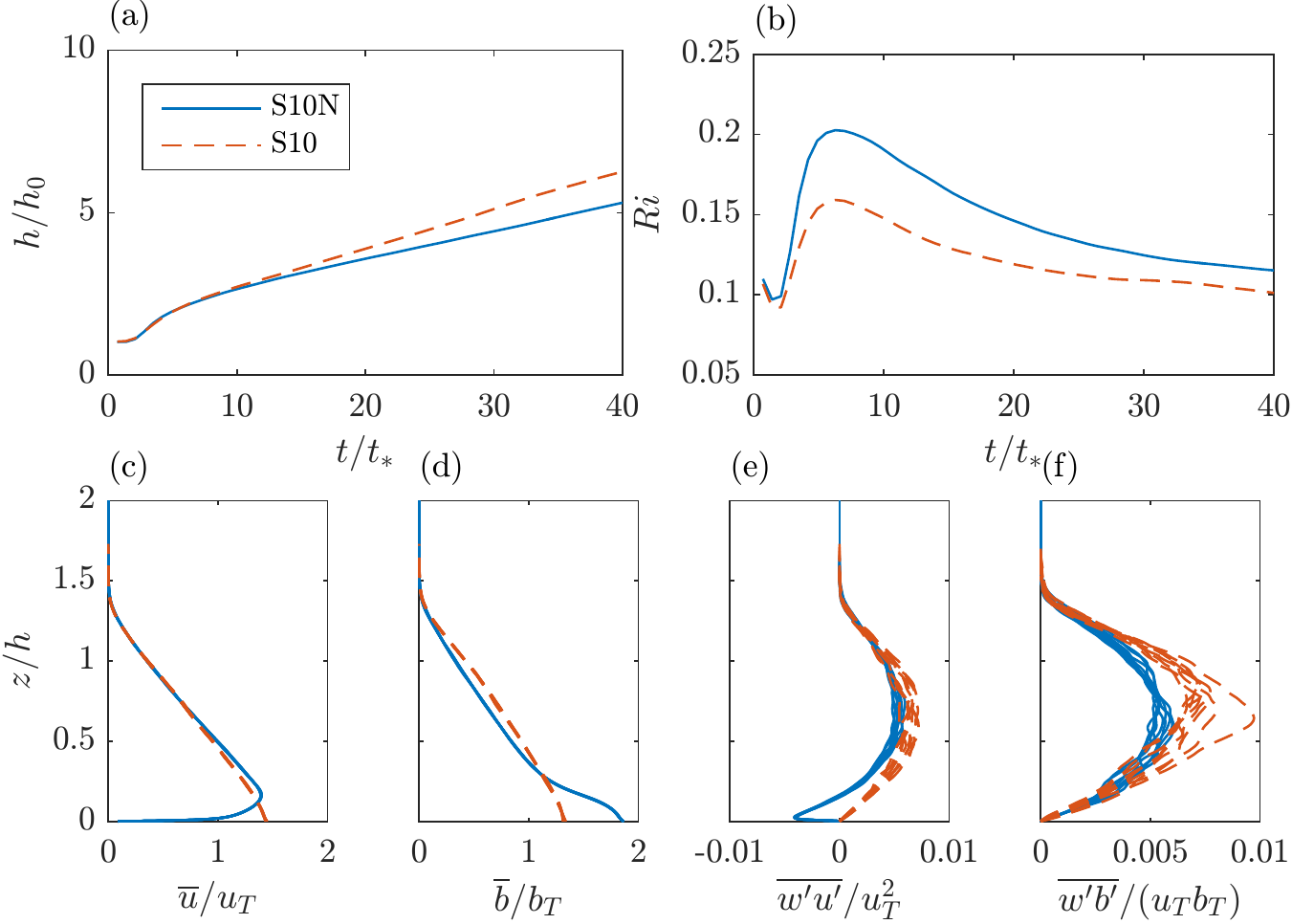}
\caption{Time variation of: a) layer thickness $h$;
 and bulk Richardson number $\Ri$, as defined in (\ref{eq:reridef}).
Scaled wall-normal variation  of scaled and spatially-averaged: c) $\overline{u}$;
 d) $\overline{b}$; e) $\overline{u' w'}$; and  f) $\overline{w' b'}$
 at
time intervals $t/t_* = 0.7$.
Note that the multiple
profiles for $\overline{u}$ and $\overline{b}$ show strong
self-similarity, are approximately linear in particular away from the
wall, and
extend beyond $z/h=1$. Results for simulations S10N and S10 are
shown with solid blue and dashed red lines respectively.}
\label{fig:NSFS_SS}
\end{figure}

The evolution of $h$ and $\Ri$ are shown in figures  \ref{fig:NSFS_SS}(a,b).
The layer thickness $h$ grows a little faster for S10 than for S10N,
indicating, perhaps unsurprisingly, slightly higher entrainment for
flows
with no-slip boundary conditions.
The evolution of $\Ri$ in time shows an initial growth to a maximum,
after which $\Ri$ reduces monotonically and approaches a steady state
for $t/t^*>30$; the final value of $\Ri$ is once again slightly higher
for simulation S10N than for simulation S10.

The approach to a constant value of $\Ri$ is qualitatively similar to
the
behaviour observed for turbulent plumes,  where plumes
attain a constant value
of the appropriately defined $\Ri$ far away from the source,
i.e. they attain `pure plume balance'. If the flow has an excess of
momentum at the source the plume is referred to as being forced
\citep{Morton1973},
while if it has a deficiency of
momentum flux it is referred to as being lazy \citep{Hunt2001},
although it has also  been interpreted in a geophysical context as
rising from a too large or `distributed' source \citep{Caulfield1995}.
In either case, the pure plume balance solution is a global attractor in an
unstratified
ambient \citep{Caulfield1998}, and   the flow inevitably adjusts itself
due to the work done by gravity until it attains pure plume balance.
Restricting attention to $t/t_*>7$ (when the flow has transitioned to
turbulence), the gravity current is `lazy', approaching `pure'
behaviour for $t/t^*$ greater than (approximately) 30.

The (strong) self-similarity of the spatially-averaged velocity
$\av{u}$ and buoyancy $\av{b}$ profiles are shown in figures
\ref{fig:NSFS_SS}(c,d), although,
unsurprisingly, the solutions are very different near the wall in the
`inner layer'.
 In particular, the `toe' observed in the buoyancy profile for S10N is
caused by the fact that the turbulence shear production $P_S$ defined
as
\begin{equation}
P_S=-\overline{w'u'} \frac{\partial \overline{u} } {\partial
  z}, \label{eq:psdef}
\end{equation}
 is zero at the
velocity maximum and that as a consequence turbulence levels are
low. This implies that buoyancy is `trapped' near the (smooth) wall for
flows with no-slip boundary conditions,  an effect reported by \cite{Ellison1959}.
Importantly though, the flow profiles sufficiently far from the wall
are very similar for the two simulations.
Figures \ref{fig:NSFS_SS}(e,f) show the wall normal transport of the
streamwise momentum and buoyancy, respectively.
It is apparent that there is a large negative momentum flux for
simulation S10N, associated with the particular near wall dynamics in
the inner layer,
and in general,
the free-slip fluxes are slightly larger than the no-slip fluxes,
while there is not such a strong collapse as for the $\av{u}$ and
$\av{b}$
profiles.

The fundamental budget of interest for mixing is the budget of turbulence kinetic energy $e$, which for the temporal gravity current is given by
\begin{equation}
    \label{eq:tke}
  \frac{\partial e}{\partial t} =
 \underbrace{-\overline{w'u'} \frac{\partial \av{u}}{\partial z}}_{P_S}\ \
 \underbrace{-\overline{u'b'} \sin \alpha + \overline{w'b'} \cos \alpha}_{P_B}
- \varepsilon.
\end{equation}
where $P_S$ is the shear production and $P_B$ is the buoyancy
production of turbulent kinetic energy with dissipation rate
$\varepsilon = \nu \overline{(\partial u_i' / \partial x_j)^2}$.
Since we are considering a flow upon a slope, there is a choice to be made as to the definition of the important mixing parameters known as
the flux Richardson number $\Ri_f$, the turbulent flux coefficient $\Gamma$ \citep{osborn80} and the gradient Richardson number $\Ri_g$.
We define these quantities as follows:
\begin{equation}
 \quad \Ri_f(z,t) =
\frac{-P_B}{ P_S} , \quad\quad
 \Gamma(z,t)=\frac{-P_B}{\varepsilon}, \quad\quad
 \Ri_g(z,t) = \frac{{\partial \overline{b}}/{\partial z} \cos
  \alpha}{\left ({\partial \overline{u}}/{\partial z} \right )^2 }.
\label{eq:rigrifgamdef}
\end{equation}
For  the gradient Richardson number $\Ri_g$ we have chosen to use the component of the density gradient
in the direction of gravity in the numerator, which
is a self-consistent generalisation of
the conventional interpretation of a gradient Richardson number as a quantification of the relative importance of destabilising shear to stabilising density gradients.

For the two (potentially related) flux quantities
$\Ri_f$ and $\Gamma$ we have chosen to use the
full buoyancy production of turbulent kinetic energy, which for
non-zero
$\alpha$ \rev{(as always considered here)} has in general a non-zero along-slope component.
This choice means that the energy-based
interpretation of these terms must be done with care, particularly
for larger values of $\alpha$.
Indeed, for
sufficiently steep angles, it is to be expected that
the buoyancy production term will change sign and become negative as the dense flow
down the slope will actually lead to an increase
in the turbulent kinetic energy, rather than a decrease (ultimately
due to an exchange into the potential energy reservoir).
It is important to appreciate that the quantities in (\ref{eq:rigrifgamdef}) are in general functions of $z$ and $t$, involving as they do only
horizontal averages, and so there is not necessarily a simple way
to relate $\Gamma$ and $\Ri_f$,
\rev{or indeed to compare with other studies
  of related flows, such as the experimental flows with the same slope
  angle discussed in \cite{Odier2014}. }
(See \cite{Wells2010} for a further
discussion of this subtle, and often not appreciated point.)

\begin{figure}
\centering
\includegraphics{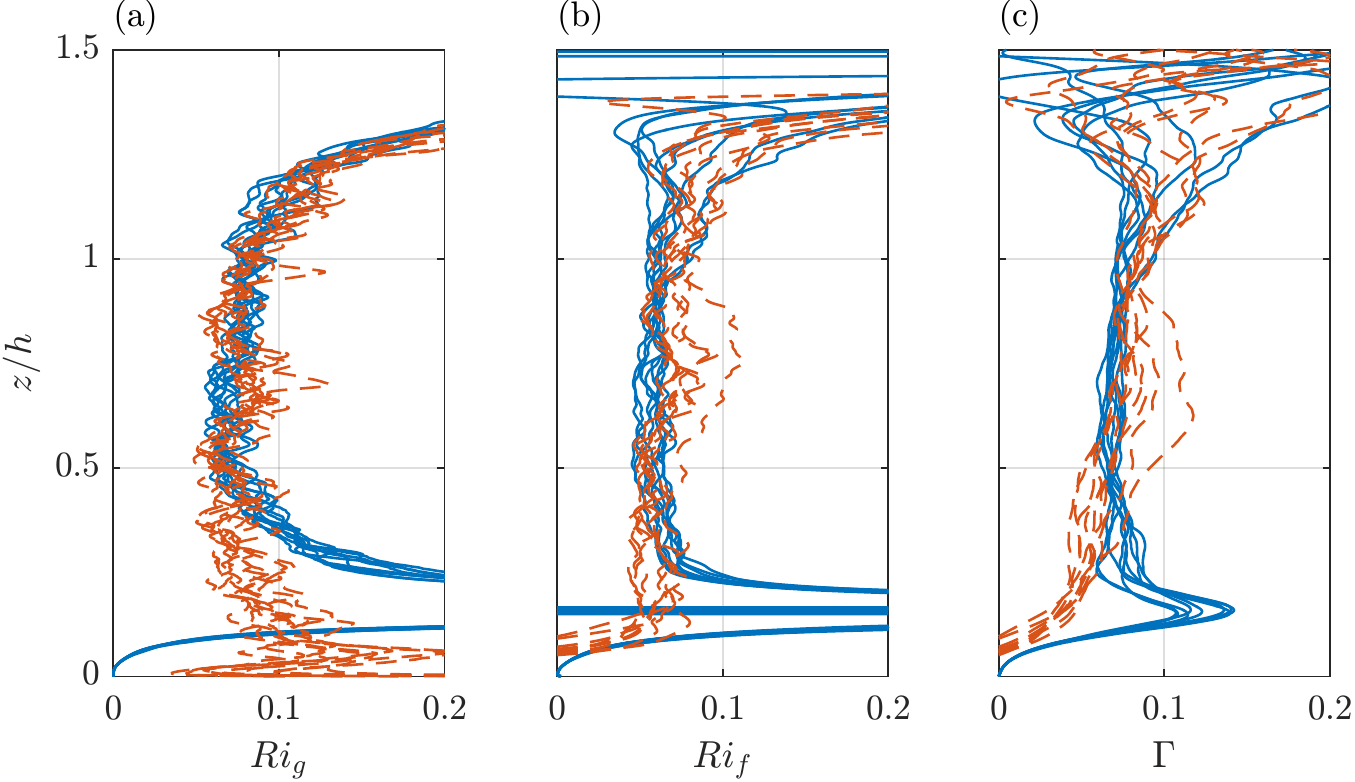}
\caption{Scaled wall-normal variation (\rev{over time interval documented in Table \ref{tab:simdata}}) of: (a)
  Gradient Richardson number $\Ri_g$; (b) Flux Richardson number
  $\Ri_f$; and (c) turbulent flux coefficient $\Gamma$ as defined
in (\ref{eq:rigrifgamdef}) for simulation S10N (red lines) and
simulation S10 (blue lines).}
\label{fig:NSFS_Ri}
\end{figure}

For the temporal gravity current, both the mean and turbulence kinetic energy are expected to scale with $B_0$ and are therefore independent of time in the fully self-similar regime.
The turbulence kinetic energy budget will then be in a statistically steady state, and thus
the two production  terms when appropriately integrated in the wall-normal direction must balance the wall-normal-integrated dissipation rate:
\begin{equation}
\mathcal{P}_S + \mathcal{P}_B = \mathcal{E}
\label{eq:tkebud_integral}
\end{equation}
Here, the caligraphic scripts denote quantities integrated over the wall-normal direction, i.e.\ $\mathcal{P}_S = \int_0^\infty P_S \d z$, $\mathcal{P}_B = \int_0^\infty P_B \d z$ and $\mathcal{E} = \int_0^\infty \varepsilon \d z$. Note that this implies that
\begin{equation}
\langle \Ri_f \rangle\simeq
\frac{\langle\Gamma\rangle}{1+\langle\Gamma\rangle},
\label{eq:rifgamma}
\end{equation}
where the angle brackets denote quantities calculated from the integral properties.
%
The quantity $\langle \Ri_f \rangle$ quantifies the proportion of the energy injected into the flow which has led to increases in the potential energy through
irreversible mixing processes, and $\mathcal{P}_B$ is expected
to be a sink of kinetic energy ultimately leading to
an increase in the flow potential energy, although particularly
for flows where the Boussinesq approximation does not apply, the
energy pathways are more convoluted  \citep[see][for detailed discussions]{Tailleux2013,Scotti2015}.
Understanding and parameterising such partitioning is a central
challenge in oceanography \rev{\citep[see][for reviews]{Ivey2008,Ferrari2009}}.

It is apparent from figure \ref{fig:NSFS_Ri} that  the values of these three quantities are quite similar
for the two simulations with different wall boundary conditions yet small angle $\alpha=10^\circ$, except in the near wall inner layer $z/h
\lesssim 0.3$, and indeed are quite close to constant over $0.3 < z/h
< 1.0$.
The evidently self-similar (and essentially linear) distributions of
$\overline{u}$ and
$\overline{b}$ naturally lead to constant values
of the gradient Richardson number $Ri_g$ over much of the
(self-similar) depth of the current. The perhaps more surprising observation
is that
the flux Richardson number $\Ri_f$ and the (apparently
related) turbulent flux coefficient $\Gamma$ are close to constant
across the interior of the current for $0.3 < z/h < 1.0$.
Indeed, although as already noted there is no necessity for the simple
relationship in (\ref{eq:rifgamma}) to hold, $\Gamma$ is indeed slightly more than $Ri_f$.

There is no immediately obvious reason why $\Ri_f(z,t)$ (and $\Gamma(z,t)$)
should be approximately
constant and similar to $\Ri_g(z,t)$, but it is entirely consistent
with the behaviour of the flows in stratified plane Couette flow,
where \cite{Deusebio2015} found a close to linear relationship
between appropriate definitions of
a gradient Richardson number and flux Richardson number, which applies
at all heights within the flow.
Furthermore,  this is also evidence (as discussed further below) that
the turbulent Prandtl number for this flow is an $O(1)$ constant.
The turbulent diffusivities for momentum and buoyancy, (once
again in general functions of $z$ and $t$) and the
associated turbulent Prandtl number are defined as
\begin{equation}
\label{eq:GDH}
K_m = -\overline{w'u'} \left( \frac{\partial \overline{u}}{\partial z} \right)^{-1}, \quad
K_\rho = -\overline{w'b'} \left( \frac{\partial \overline{b}}{\partial
    z} \right)^{-1}, \quad
Pr_T= \frac{K_m}{K_\rho},
\end{equation}
and so it is straightforward to establish
that
\begin{equation}
\Ri_f  = \frac{\Ri_g}{Pr_T} \left [ 1 -
  \frac{\overline{u'b'}}{\overline{w'b'}} \tan \alpha \right ]
.\label{eq:rifrigrelate}
\end{equation}
Since the angle $\alpha$ is relatively small, the term in the square
bracket is close to unity  and so the observation that the \textit{flux} Richardson number $\Ri_f$ is close to constant and similar in magnitude to the constant \textit{gradient} Richardson number $\Ri_g$ implies
that $Pr_T \sim O(1)$ and close to constant (with $z$ and $t$).
We observe that for all heights in  the current sufficiently far away from the wall,  the flow `chooses' a balanced coupled configuration for which $\Ri_f$ and $\Gamma$ are constant across the outer  layer.
This \rev{inherently emergent dynamical behaviour  of spatially-constant $\Ri_f$
  and $\Gamma$ occurs}
for all simulations, although $\Ri_f$ and  $\Gamma$ do have a  non-trivial dependency on $\alpha$ as discussed  further below, not least because of the increasing influence of the  along-slope term in $P_B$ as the slope angle increases.

\subsection{Dependence on the slope angle}

Having established that the boundary conditions have  relatively
limited influence on turbulent entrainment,
\rev{at least for such purely buoyancy-driven down-slope gravity currents,}
we restrict attention to
free-slip velocity boundary conditions so that by construction we can
focus on outer layer dynamics.  We consider simulations with a wide
variety of  angles $\alpha$ between 2 and 90 degrees.
It is important to appreciate that  the simulation
with $\alpha=90^{\circ}$ closely resembles the flow of a  temporal
plume,
and is in a very real sense a qualitatively different flow than a temporal gravity current.

\begin{figure}
\centering
\hspace{-4mm}
\includegraphics{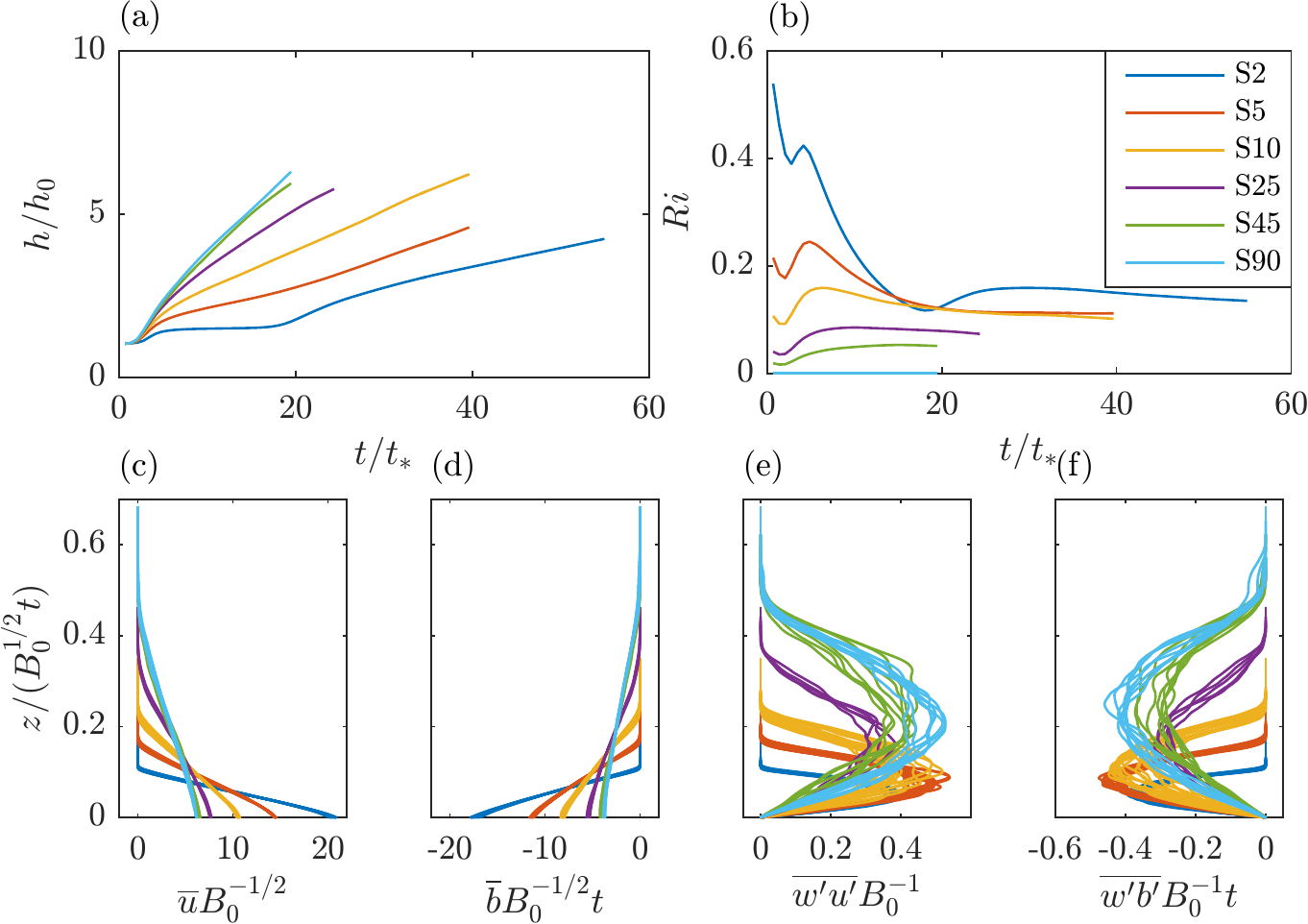}
\caption{Properties of the results of free-slip simulations
for a range of slope angles, as marked with different line types.
Time variation of: (a) $h$; and (b) $\Ri$.
Scaled wall-normal variation of
spatially-averaged: (c) $\overline{u}$;
 (d) $\overline{b}$; (e) $\overline{u' w'}$; and  (f) $\overline{w'
   b'}$ \rev{over the time interval documented in Table \ref{tab:simdata}}.
}
\label{fig:alpha_SS}

\end{figure}

Figure \ref{fig:alpha_SS}(a) shows the time evolution of the current
depth $h$ for simulations with a range of angles,  showing that the
depth grows approximately linearly with time as expected.
It is also clear that $h$ grows more rapidly as the angle
$\alpha$ is increased.
As shown in figure \ref{fig:alpha_SS}(b),
the bulk Richardson number $\Ri$
approaches a constant value once the flow is self-similar, with the
asymptotic
value decreasing with increasing angle.

The extents of the self-similarity of various spatially-averaged
characteristics
of the flows are shown in figures
\ref{fig:alpha_SS}(c-f). In
contrast to figure
 \ref{fig:NSFS_SS},
$B_0$ and $t$ are used to scale
both the wall-normal
coordinate  and the various flow characteristics
instead of $u_T$ and $b_T$, so that
it is possible to identify
the differences in absolute amplitudes between simulations
with relatively small angles (i.e. $\alpha=2^\circ,5^\circ, 10^\circ$)
and relatively larger angles (i.e. $\alpha=25^\circ,45^\circ,
90^\circ$).
Clearly, the currents from simulations with relatively low angles
are shallower  and have higher amplitudes of $\overline{u}$ and
$\overline{b}$
compared to the currents along relatively steeper slopes.
Flows along slopes with small $\alpha$ require
more work to be done for an increase in potential energy,
and so the flow
will accelerate until this can be achieved,
and the self-similarity is at least approximately achieved.
As is apparent in figures \ref{fig:alpha_SS}(e,f),  the turbulent fluxes
are also close to self-similar, and it is striking that the maximum amplitude is similar for all values of $\alpha$, although the location of the maximum does depend on $\alpha$, with the maximum being closer to the wall for flows with smaller slope angle $\alpha$.

\begin{figure}
\centering
\includegraphics{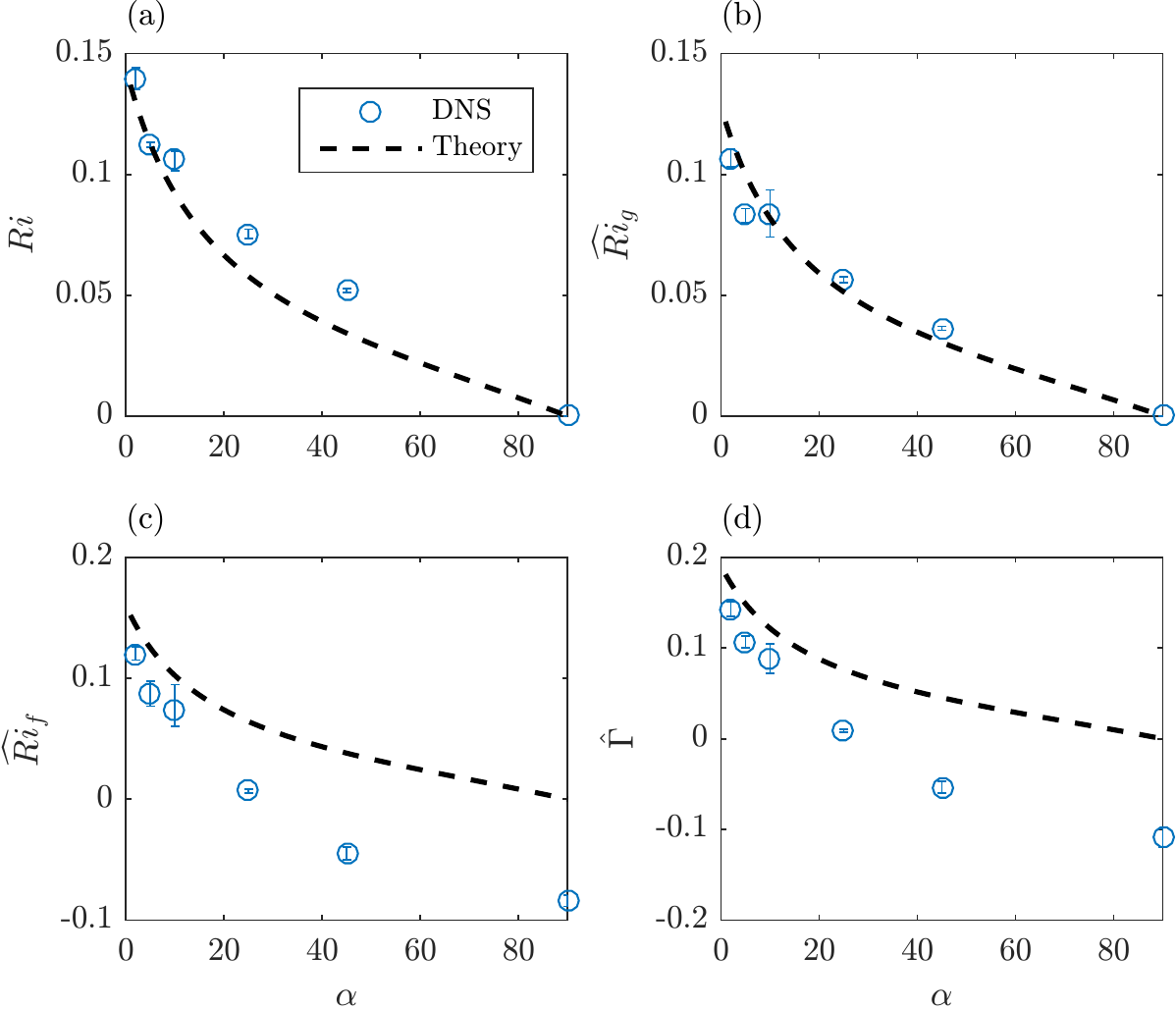}
\caption{Variation with slope angle
$\alpha$ of the numerical data (marked with circles) and theoretical predictions (marked with dashed lines)
 for:
(a) $\Ri$, modelled by (\ref{eq:ririg}); 
(b) $\widehat{\Ri_g}$, modelled by (\ref{eq:ririg});
 (c) $\hat{\Ri_f}$, modelled by (\ref{eq:rifgam});
and (d) $\hat{\Gamma}$, modelled by (\ref{eq:rifgam}).
A hat denotes an average over the outer layer interval $z/h=[1/2,1]$.
}
\label{fig:alpha_alphadep}

\end{figure}

In Figure \ref{fig:alpha_alphadep}, the observed dependence of the mixing parameters on the slope angle $\alpha$ is presented by circles.
Here, the hatted quantities denote average values over the outer layer interval $z/h = [1/2, 1]$ (see Figure \ref{fig:NSFS_Ri}).
The error bars show the difference between the largest and smallest value within the time window $t_{stat}$ over which statistics were gathered (the statistics were gathered at intervals $t/t_*=0.7$).
Shown with the dashed line are results from a theoretical model which will be introduced in \S \ref{sec:approx}.
Clearly, all indicators display similar behaviour in the sense that their values increase as the slope angle reduces.
Importantly, though, the values remain bounded and well below the
critical linear stability threshold of $\Ri=1/4$.
$\widehat{Ri}_f$ and $\hat{\Gamma}$ are positive only for the three
low-angle \rev{(but still finite slope)} cases.
For $\alpha > 25^o$, the along-slope turbulence kinetic energy contribution of $P_B$ becomes larger than the perpendicular contribution, and thus both $\widehat{Ri}_f$ and $\hat{\Gamma}$ will become negative.
As already discussed, in such a regime, the flux Richardson number
$\Ri_f$ and the turbulent flux coefficient $\Gamma$ are not appropriately defined.

\subsection{Turbulent entrainment}

\begin{figure}
\centering
\includegraphics{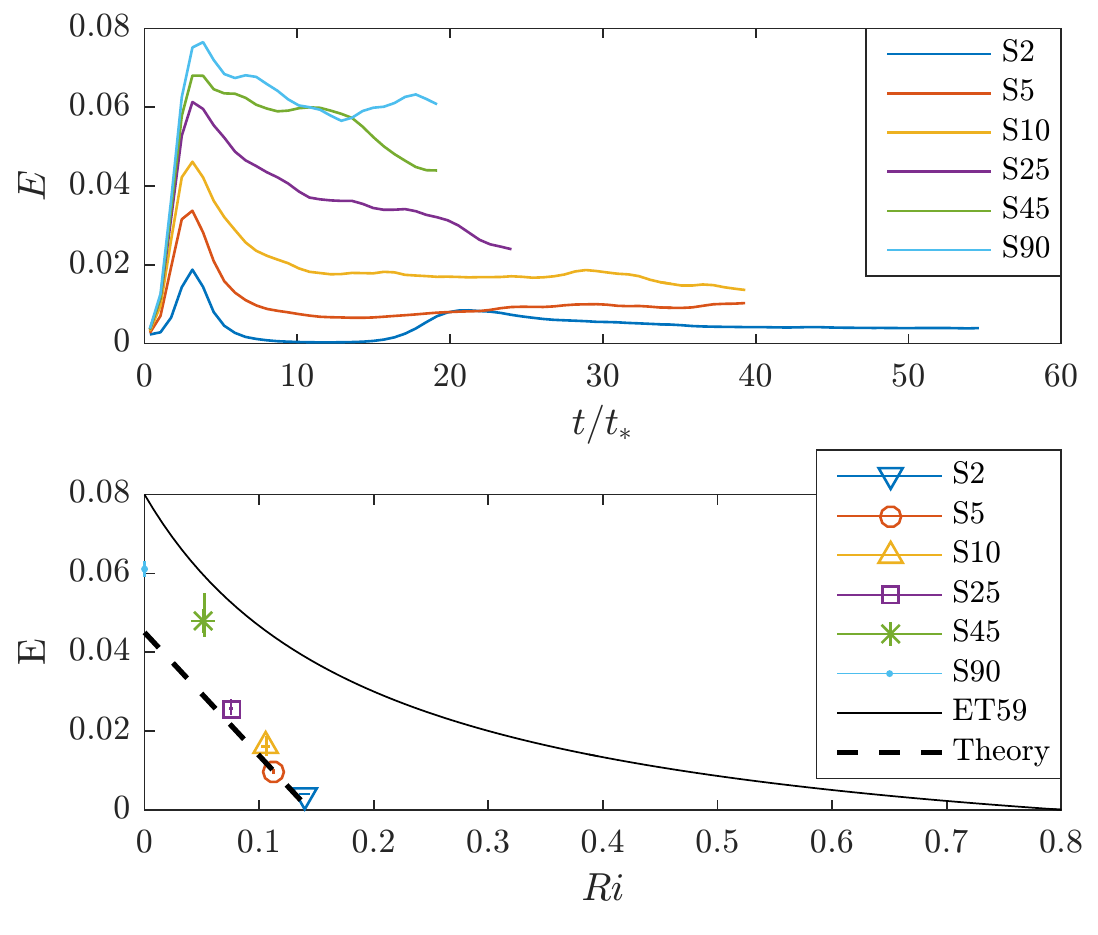}
\caption{(a) Variation with time of the entrainment parameter $E$ as
defined in (\ref{eq:edef}) for various simulations, plotted
with different line types.
(b) Variation with $\Ri$ of the averaged
entrainment parameter $E$ as
defined in (\ref{eq:edef}) for various simulations, plotted
with different symbols. The classical empirical parameterisation ET59 as defined in
(\ref{eq:et59})
is plotted with a solid line, while the new proposed parameterisation
(\ref{eq:eparam}) is plotted with a dashed line. }

\label{fig:alpha_E}
\end{figure}

Turbulent entrainment, as quantified by the entrainment parameter \citep{vanReeuwijk2017}
\begin{equation}
\label{eq:edef}
E  = \frac{1}{u_T} \frac{\d h}{\d t},
\end{equation}
is naturally closely related to turbulent mixing.
In figure \ref{fig:alpha_E}(a), we plot the temporal variation of
$E$ for all six simulations.
 Clearly, $E$ increases with  $\alpha$. Since $\alpha$ is directly
 related to the bulk Richardson number $\Ri$ through its definition
(\ref{eq:reridef}), this relationship suggests, analogously to the classic
parameterisation of Ellison \& Turner discussed in \S \ref{sec:intro},
that  $E$ has a functional dependence on  $\Ri$.
Despite keeping $B_0$ the same for all simulations,
quantifying $E$ in the simulations at the  smallest \rev{(yet
  crucially still finite, as noted above)} angles is
challenging since
there is a lower turbulence intensity (not shown)
as a result of the (relatively) stronger stratification.
This is particularly evident in the data extracted from simulation S2,
in which the flow  almost completely relaminarises
after the initial burst associated with
Kelvin-Helmholtz-like shear instabilities as the flow accelerates
from its initial condition.
However,
the very low, \rev{but inherently transient,} level of turbulence (evident for $t/t_* \simeq 10$)
implies that the fluid layer will accelerate due to the low level of
dissipation until a second transition occurs  after which  `pure'
behaviour emerges for $t/t_*>40$. \rev{This second transition appears
  to lead to a quasi-stationary turbulent state. Therefore, it is not
  appropriate
  to interpret this flow  as exhibiting `intermittent' turbulence,
  but rather that our chosen initial conditions are in some sense
  imbalanced, and so there is a relatively long initial
  transient evolution, exhibiting non-monotonic turbulence
  intensity, until the flow self-organises into its eventual
  quasi-stationary turbulent state.}

The entrainment law $E(\Ri)$ is shown in the bottom plot of figure
\ref{fig:alpha_E}, together with the Ellison and Turner entrainment
law defined in (\ref{eq:et59}).
Here, $\Ri$ is the value associated with the `pure' gravity current
after initial transients have decayed, and the velocity and
buoyancy distributions have evolved into the clearly coupled (and close
to linear) distributions as shown in figure \ref{fig:alpha_SS}(c).
Precisely, the value of $E$ was calculated by averaging over the
interval $t_{stat}$ (table \ref{tab:simdata}), and the horizontal and
vertical error bars denote the maximum and minimum observed values of
$\Ri$ and $E$ over this time interval.

Interestingly,  $\Ri$ remains constrained between $0 < \Ri < 0.2$ for all angles.
Crucially, when defined in this way, we observe that it is impossible
to reach the higher values of $\Ri$ observed in the \cite{Ellison1959}
experiments: the coupled velocity and buoyancy distributions
inevitably force the Richardson number to be bounded above by $0.2$
when there is any observable turbulence at all, consistently
with the recent observations of  \cite{Deusebio2015} and
\cite{Zhou2017} in plane Couette flow.
Furthermore, the values of $E$ as quantified from our numerical
simulations are  lower than those reported experimentally
by \cite{Ellison1959} \rev{and \cite{Odier2014} although direct
  quantitative comparison is challenging, not
  least due to the different forcing mechanisms of the different
  flows.}

Of course, the spatially evolving experimental  flow geometry is qualitatively different from the inherently temporal flow geometry considered here.
 It is also at least plausible that the gravity current in their
 experiments was not yet `pure' in a statistically steady state, which would go some way to explain their higher observed values for $\Ri$ and $E$.
Furthermore, as noted in \S \ref{sec:intro}, it is also conceivable
that their experimental data were contaminated by end-effect
mixing at a `head' being swept back along the current,
as such mixing is expected to be substantially stronger, as
investigated
in detail by \cite{Sher2015}.
Whatever the reason for the mismatch, it is apparent that there
is a marked decrease in $E$ with $\Ri$, and there
is a clear `switch-off' of entrainment (quantified in this way) at a maximum value of $\Ri$.

Focusing on the simulations S2, S5 and S10 with relatively
small slope angles, (where the buoyancy production $P_b < 0$, and
so the expected sink of turbulent kinetic energy due to
mixing occurs)  the entrainment law can be reasonably well
approximated by a linear relation of the form
\begin{equation}
E=a (\Ri_{\max} - \Ri); \ a=0.31,  \ \Ri_{\max} = 0.15,
\label{eq:eparam}
\end{equation}
 as marked in  figure
\ref{fig:alpha_E}(b) by the dashed line.
We present a detailed
theoretical justification for this model in \S \ref{sec:approx}, but
we once again stress that for these simulations, the mean velocity and
buoyancy profile are both close to linear and appear to be 
coupled in a way which strongly suggests that there is a maximum value
of $\Ri$ beyond which turbulence cannot be sustained, \rev{at least in
  the flows considered here, purely driven by buoyancy forces
  down slopes of small (but still finite) angle.}

\section{Model development}
\label{sec:model}

\subsection{Turbulence parameterisation}
\label{sec:turb}

As demonstrated in \cite{Krug2017}, the vertical length scale is
proportional to $\hat{e}^{1/2} / \hat{S}$, which implies that the
turbulence is `shear-dominated' \citep{Mater2014}.
It is thus natural to investigate whether a turbulence
parameterisation
can be constructed using the strain rate
$S = |\partial \overline{u} / \partial z|$
and the turbulence kinetic energy $e$ to scale the
eddy diffusivity calculated from three simulations with relatively
small slope angles, i.e. the simulations S2, S5 and S10.
For the turbulent diffusivities defined in \eqref{eq:GDH}, this suggests that
\begin{equation}
K_m = c_m e/S, \quad\quad K_\rho = c_\rho e/S. \label{eq:kmparam}
\end{equation}

\begin{figure}
\centering
\includegraphics{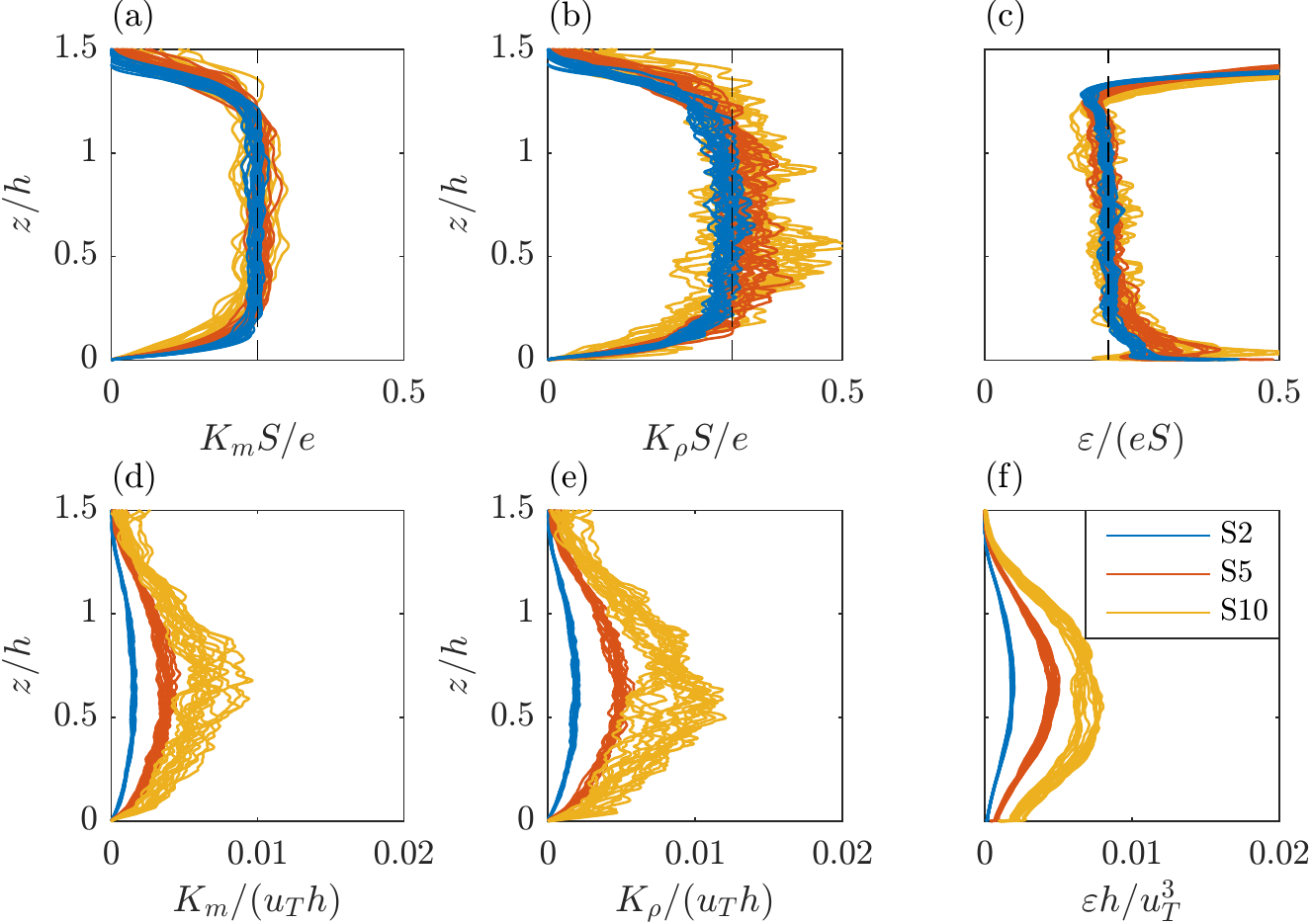}
\caption{Variation with scaled height $z/h$ for the three
  simulations
S2 (blue lines),
S5 (red lines)
and S10 (yellow lines) at
several time instants (\rev{see Table \ref{tab:simdata} for time interval})
for:
(a)  $K_m S/e$;
(b)  $K_\rho S/e$;
(c)  $\varepsilon / (eS)$;
(d) $K_m/(u_T h)$;
(e) $K_\rho/(u_T h)$;
(f) $\varepsilon h/u_T^3$.
In panels (a-c) there is clear collapse, and the
relevant empirical constants of proportionality $c_m=0.25$, $c_\rho=0.31$ and $c_{\varepsilon}=0.21$ are marked with vertical dashed lines.
}
\label{fig:alpha_nuT}
\end{figure}

In figure \ref{fig:alpha_nuT}(a), we plot
$K_m S/e$ for all three \rev{small-slope} simulations as a function of $z/h$ at
a range of times. The vertical dashed line shows $c_m=0.25$,
demonstrating that this parameterisation is an excellent fit for a
wide range of times for all three simulations, particularly 
sufficiently far away from the wall, i.e.\ $0.3 < z/h < 1$.
Conversely, in figure \ref{fig:alpha_nuT}(d) by plotting
$K_m/(u_T h)$, we demonstrate that there is no collapse when scaled
with the similarity variables.
Similarly, for the turbulent scalar diffusion $K_\rho$, we observe a
collapse upon plotting $K_\rho S / e$, as shown in figure
\ref{fig:alpha_nuT}(b), with an approximately constant value
sufficiently far away from the wall  of $c_\rho = 0.31$ (as plotted
with a vertical dashed line).
This implies that for three lowest angle cases, the turbulent Prandtl
number $\Pr_T = K_m / K_\rho = c_m / c_\rho = 0.8$ \emph{independent}
of the flow angle, thus clearly suggesting that $Pr_T$ remains finite,
\rev{and indeed is of $O(1)$}.
Once more there is no collapse upon scaling $K_m$ by $u_T h$, as shown
in figure \ref{fig:alpha_nuT}(e).

Similarly, we predict that the dissipation rate  $\varepsilon$ can also be scaled appropriately with $e$ and the strain rate $S$, which on dimensional grounds implies that  \begin{equation}
\label{eq:turbparam2}
\varepsilon = c_\varepsilon e S,
\end{equation}
for some constant $c_\varepsilon$. We plot
$\varepsilon/(e S)$ in figure \ref{fig:alpha_nuT}(c), and once again
there is
a clear collapse, with an approximate value of
 $c_\varepsilon = 0.21$ away from the immediate vicinity of the wall, as marked
 on the figure with a dashed line. Conversely, as shown in figure
 \ref{fig:alpha_nuT}(f), there is no collapse of $\varepsilon$
when scaling with $u_T^3/h$.

It is also interesting to note, using the definitions of
$\Ri_f$, $\Gamma$, $P_S$ (\ref{eq:rigrifgamdef}), and $K_m$ (\ref{eq:GDH}), as well as the (verified)
parameterisation (\ref{eq:kmparam}),
that
\begin{equation}
\varepsilon= \left (\frac{\Ri_f}{\Gamma} \right ) P_S =
\left (\frac{\Ri_f}{\Gamma} \right ) K_m S^2 \quad \Rightarrow \quad
\varepsilon =\underbrace{\left (\frac{\widehat{\Ri}_f}{\hat{\Gamma}} \right ) c_m}_{c_\varepsilon} e S. \label{eq:cnuces}
\end{equation}
Comparing this relationship
to the empirical observation that
$c_\varepsilon=0.21$ and $c_m=0.25$, this suggests $\widehat{\Ri}_f / \hat{\Gamma} = 0.84$ and thus
that $\widehat{\Ri}_f  < \hat{\Gamma}$, consistently with (\ref{eq:rifgamma}).

\begin{figure}
\centering
\includegraphics{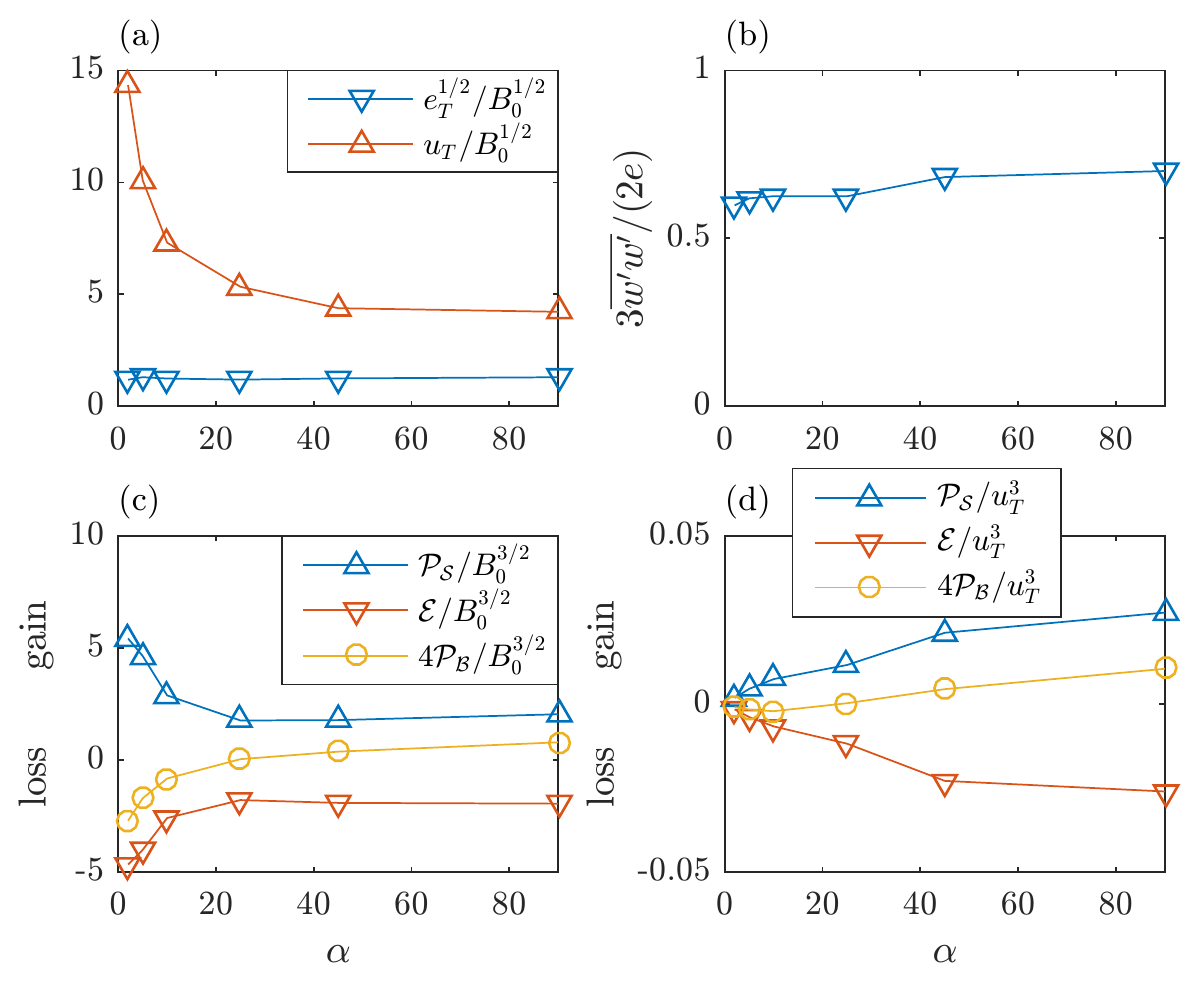}
\caption{Variation with slope angle of (a) the characteristic velocities $u_T$ and turbulence $e_T^{1/2}$, (b) the anisotropy of the turbulence and (c,d) the vertically integrated terms in the turbulent kinetic energy equation: shear
production $\mathcal{P}_S$ (marked with blue triangles);
dissipation rate $\mathcal{E}$ (red inverted triangles);
buoyancy production $4\mathcal{P}_B$ (yellow circles).
In panel (c) these quantities
are scaled with $(B_0^3)^{1/2}$,
while
in panel (d) they are scaled with $u_T^3$.
}
\label{fig:alpha_tkebud}
\end{figure}

It is apparent that for the key properties of the turbulence in
this flow, the appropriate scalings are  not based around the conventional integral velocity scale $u_T$ and $h$.
This is clearly demonstrated in figure \ref{fig:alpha_tkebud}(a), where $u_T$ and $e_T^{1/2}$ are displayed as a function of the slope angle $\alpha$.
As mentioned earlier, these quantities are both expected (and observed) to scale according to $B_0^{1/2}$ in the self-similar regime.
It is striking that $e_T$ is virtually independent of $\alpha$, but
$u_T$ increases dramatically as the slope angle reduces. \rev{Indeed,
  the structure of the $u_T$ variation is consistent with $u_T$
  diverging
as $\alpha \rightarrow 0$. As is apparent from the definition
\eqref{eq:b0def} of the integral forcing $B_0$, the forcing buoyancy
$b_0 h_0$ must diverge as $\alpha \rightarrow 0$ to ensure the
distinguished limit that $B_0$ remains constant, and so it is perhaps
unsurprising that $u_T$ becomes very large as $\alpha$ approaches
small values.}
The \rev{physical} picture that emerges is that the flows with higher Richardson
numbers for lower slope angles $\alpha$ require higher shear (and thus higher $u_T$) to sustain a similar level of turbulence.
This will be explained in detail by the conceptual model we develop in the next section.
What will not be revealed by the model is why the turbulence level
$e_T/B_0^{1/2}$ is virtually independent of $\alpha$.
In figure \ref{fig:alpha_tkebud}(b), we plot the variation of $(3/2) (\overline{w'
  w'}/e)$ with slope angle $\alpha$ to quantify the extent of
anisotropy within the flow. Unsurprisingly, due to the inherent
greater magnitude of the streamwise velocity $u$, this quantity is
always less than one, and exhibits only a slight dependence on angle 
when scaled in this way, decreasing slightly as $\alpha$ \rev{tends to
  small values,}
due naturally to the stabilising effect of stratification becoming
somewhat more significant. 
However, there are no signs of a transition to strongly stratified
turbulence, and this cannot be an explanation for the observation of
$E \rightarrow 0$ as \rev{$\alpha$ tends towards zero.}

In figure \ref{fig:alpha_tkebud}(c) and (d), we plot  the variation with
slope angle $\alpha$ of the three key terms $\mathcal{P}_S $, $4\mathcal{P}_B$ (the prefactor is included for visual purposes only) and $\mathcal{E}$ in the  integral turbulent kinetic energy equation \ref{eq:tkebud_integral} and averaged over the time interval $t_{stat}$ (Table \ref{tab:simdata}).
In figure \ref{fig:alpha_tkebud}(c), the various terms are
scaled using $(B_0^3)^{-1/2}$, which reveals the appropriate and
expected dynamics for small angles, showing that the
actual production of turbulence increases. This can be understood from
the fact that the velocity gradient is much larger (see figure
\ref{fig:alpha_SS}) as the flow will keep accelerating until an
essentially equilibrium state is attained.
As mentioned earlier, the buoyancy production term $\mathcal{P}_B$ is negative for the smallest three
angles and changes sign around $\alpha = 20^o$, which can be understood in terms of the relative size of the two terms in $\mathcal{P}_B$ (\ref{eq:tke}), with the contribution of the first (generally positive) term dominating for large $\alpha$ and the second (generally negative) term for small $\alpha$.
Clearly, not least because of the opposing effects of these
two terms, the main overall balance is between shear production $\mathcal{P}_S$ and dissipation $\mathcal{E}$, and  $\mathcal{P}_B$ only makes a relatively small contribution to the overall budget.

\rev{Although it is apparent from figure \ref{fig:alpha_tkebud}(c)
  that the magnitude of the various key terms in the integral turbulent kinetic energy
  equation \ref{eq:tkebud_integral} increase as $\alpha$ decreases,
  their rate of increase
  is clearly significantly smaller than that exhibited by $u_T$ as
  shown in figure \ref{fig:alpha_tkebud}(a).
  Figure \ref{fig:alpha_tkebud}(d)} shows once more the integral turbulence kinetic energy budget but this time scaled by $u_T^3$. Scaled in this way, as $\alpha \rightarrow 0$  the
production of turbulent kinetic energy appears  to go to zero,
\rev{although it is more appropriate to interpret this figure as
  implying
  that the production of turbulent kinetic energy becomes effectively
  insignificant as $\alpha \rightarrow 0$.}
Significantly, even though $P_B$ increases in terms of $B_0$, the relative production tends to zero as $\alpha \rightarrow 0$. This has profound implications for the entrainment as there will be no energy available for turbulent mixing.

\subsection{An approximate integral model for small $\alpha$}
\label{sec:approx}

Since we observe
 apparently coupled and essentially linear  profiles of velocity and
buoyancy,
we now construct a model for  entrainment
based around this central observation, in an attempt
to predict the apparent `switch-off' of the entrainment, particularly
for the simulations associated with a low slope angle.
The conservation equations for mean momentum, buoyancy and turbulent kinetic energy are given by, respectively
\begin{align}
  \label{eq:u_mod}
  \frac{\partial \av{u}}{\partial t}  &= -\frac{\partial \av{\fl{w}\fl{u}}}{\partial z} - \av{b} \sin \alpha, \\
  \label{eq:b_mod}
  \frac{\partial \av{b}}{\partial t} &= -\frac{\partial \av{\fl{w}\fl{b}}}{\partial z}, \\
    \label{eq:e_mod}
  \frac{\partial e}{\partial t} &=
- \overline{w'u'} \frac{\partial \av{u}}{\partial z}
+ \overline{w'b'} \cos \alpha
- \varepsilon.
\end{align}
Here, we have neglected the term $\av{u'b'} \sin \alpha$ describing
the along-slope  turbulent production by
buoyancy in the turbulent kinetic energy equation.
This  limits the validity of the approximation to small
values of $\alpha$, as we are interested in the dynamical
behaviour of the system where the dominant effect of the buoyancy
is to extract kinetic energy due to vertical mixing associated with entrainment.
The system of equations is closed using our \rev{parameterisations}
of the eddy diffusivities of momentum and buoyancy, and the scaling of the dissipation rate
in terms of the turbulent kinetic energy and the mean shear, i.e.
\eqref{eq:kmparam} and \eqref{eq:turbparam2}.

Furthermore, since we observe
that the profiles of $u$ and $b$ appear
to be self-similar, we assume a solution of the form
\begin{equation}
\overline{u}=\underbrace{a_u B_0^{1/2}}_{u_T} f_u(\eta), \quad \quad
\overline{b}=\underbrace{\rev{-} a_b \frac{B_0}{h\sin \alpha}}_{b_T} f_b(\eta), \quad \quad
e=\underbrace{a_e B_0}_{e_T} f_e(\eta)
\label{eq:solSS}
\end{equation}
where the scaled wall-normal distance $\eta=z/h$ is the natural
similarity variable and  $a_u$,
$a_b$ and $a_e$ are coefficients that will need to be determined.
Given the relative complexity of the equations we
will not attempt to obtain closed-form solutions. Instead we employ the Von Karman-Pohlhausen method \citep{Lighthill1950, Spalding1954, Schlichting2000} to construct an approximate solution to the system of coupled PDES using an ansatz
\begin{equation}
  f_u = f_b = \frac{2}{\eta_1^2} (\eta_1 - \eta), \quad \quad
  f_e = \frac{6 \eta}{\eta_1^3} (\eta_1 - \eta), 
  \label{eq:solapprox}
\end{equation}
on the interval $\eta\in[0,\eta_1]$. These profiles are constructed such that  $\int_0^\infty f_u \d \eta = \int_0^\infty f_b \d \eta=\int_0^\infty f_e \d \eta = 1$ in order to enforce consistency with $u_T$, $b_T$ and $e_T$.

Using \eqref{eq:solSS} and  \eqref{eq:solapprox}, the key integrals
of the volume, momentum and buoyancy in the current
take the forms
\begin{align}
 \int_0^{\infty}\overline{u}dz&=B_0^{1/2} h \int_0^{\eta_1}f_ud\eta = a_u B_0^{1/2} h  ,\\
 \int_0^{\infty}\overline{u}^2dz&=B_0 h \int_0^{\eta_1}f_u^2d\eta = \frac{4}{3} a_u^2 B_0 h  ,\\
 \int_0^{\infty}\overline{b}\sin\alpha dz&=-B_0 \int_0^{\eta_1}f_b
 d\eta =-a_b B_0 .
\end{align}
From the fact that the buoyancy integral is conserved and equal to $-B_0$, it follows directly that $a_b=1$.
From the definition of $h$ (\ref{eq:integrals2}),
 it follows that $\eta_1=4/3$.
Making the further scaling assumption that
$h=a B_0^{1/2} t$, for a further empirical constant $a$, we
obtain the following expressions for the
main quantities of interest:
\begin{align}
  u_T &= \frac{M}{Q} =  a_u B_0^{1/2} , &
  E &= \frac{1}{u_T} \frac{d h}{d t} = \frac{ a }{a_u} \\
  \hat{N}^2 &=  \frac{\partial \overline{b}}{\partial z} \cos \alpha
       = \frac{9}{8}\frac{B_0}{h^2 \tan \alpha} ,&
  \hat{S} &= \left| \frac{\partial \overline{u}}{\partial z} \right| = \frac{9}{8} \frac{B_0^{1/2} a_u}{h}, \\
  \Ri &= -\frac{b_T h\cos \alpha}{u_T^2 } = \frac{1}{a_u^2 \tan \alpha} ,&
  \widehat{\Ri}_g &= \frac{\hat{N}^2}{\hat{S}^2} = \frac{8}{9}\frac{1}{a_u^2 \tan \alpha}  , \\
  \widehat{\Ri}_f &= \frac{8}{9}\frac{1}{a_u^2 \Pr_T \tan \alpha}, &
  \hat{\Gamma} &= \frac{8}{9} \frac{c_m}{c_\varepsilon \Pr_T a_u^2 \tan\alpha}.
\end{align}
\rev{Here, the hat emphasises that these quantities represent the expected average value in the outer layer.
}

Equations for the various empirical constants
can be determined by integrating (\ref{eq:u_mod}) and (\ref{eq:e_mod})
from $\eta=0$ to $\eta=\eta_1$, leading ultimately to
\begin{align}
  a a_u &= 1,\\
      \frac{9 a_u}{8} (c_m -c_\varepsilon) -\frac{c_m}{a_u \tan\alpha Pr_T} &= a .
\end{align}
Analogous integration of the
buoyancy equation  (\ref{eq:b_mod})  does not provide any further
information, confirming that
the chosen profiles are \emph{a priori} consistent with
(\ref{eq:b_mod}).   Solving for $a_u$ and $a^2$ results in
\begin{align}
a_u &= \frac{1}{a} ,&
a^2 &= \frac{9}{8} \frac{Pr_T (c_m - c_\varepsilon) \tan
  \alpha}{\tan \alpha \Pr_T+c_m} \label{eq:asquared}
\end{align}
Interestingly, the empirical constant $a_e$ remains \emph{free},
implying that these results are not directly sensitive to the actual
magnitude of the turbulent kinetic energy fluctuations.

For the mixing parameters, the model thus predicts that
\begin{align}
  \Ri &= \frac{9}{8} \frac{c_m - c_\varepsilon}{\tan \alpha +c_\rho} ,&
  \widehat{\Ri}_g &= \frac{c_m - c_\varepsilon }{\tan \alpha +c_\rho}  , \label{eq:ririg}\\
  \widehat{\Ri}_f &=  \frac{c_m - c_\varepsilon }{\tan \alpha \Pr_T+c_m}, &
  \hat{\Gamma} &= \frac{c_m }{c_\varepsilon}  \frac{ c_m - c_\varepsilon }{\tan \alpha \Pr_T+c_m}.\label{eq:rifgam}
\end{align}
Note that these expressions are fully consistent with relations
\eqref{eq:rifrigrelate}, \eqref{eq:cnuces} derived above in sections
\ref{sec:ss} and \ref{sec:turb} respectively.
Figure \ref{fig:alpha_alphadep} shows how these theoretical predictions
compare to the simulation
data.

Both $\Ri$ and $\widehat{\Ri}_g$ compare  well across  the entire range of
simulations, even -- somewhat surprisingly -- for the large angle
cases. This suggests that the along-slope buoyancy production is
not significant for these two parameters, although it
also could just be a straightforward consequence
of the fact that both $\Ri$ and $\hat{\Ri}_f$ tend to zero as $\alpha
\rightarrow 90^o$.
Conversely,  $\widehat{\Ri}_f$ and $\hat{\Gamma}$ only really compare
well to the numerical simulation data for the three relatively low-angle cases.
For $\alpha > 25^o$, the theoretical and numerically simulated values start deviating strongly, due
presumably to the omission of the along-slope turbulent kinetic energy
production due to buoyancy, as is clear from the fact that both
quantities change sign for the numerical simulation data. As already
discussed, in such a regime, the flux Richardson number $\Ri_f$ and
the turbulent flux coefficent  $\Gamma$ are not appropriately defined.

The model predicts that the maximal attainable values for the mixing
parameters occur
\rev{as $\alpha \rightarrow 0$}
and are given by
\begin{align}
  \Ri_{\max} &= \frac{9}{8} \frac{c_m - c_\varepsilon}{c_\rho} \approx 0.15 ,&
  \widehat{\Ri}_{g;\max} &= \frac{c_m - c_\varepsilon }{c_\rho} \approx 0.13  , \\
  \widehat{\Ri}_{f;\max} &=  \frac{c_m - c_\varepsilon }{c_m} \approx 0.16, &
  \hat{\Gamma}_{\max} &= \frac{ c_m - c_\varepsilon}{c_\varepsilon} \approx 0.19.
\end{align}

The entrainment parameter $E=a^2$, and the angle $\alpha$ will need
to be eliminated in order to obtain the entrainment law. This can be
achieved by using the equation for $\Ri$ in (\ref{eq:ririg}), which
can be rearranged to yield
\begin{equation}
\tan \alpha = \frac{9}{8} \frac{c_m-c_\varepsilon}{\Ri}-\frac{c_m}{\Pr_T}.
\end{equation}
Substituting this relationship into the expression for $a^2$ (\ref{eq:asquared}), it follows that the entrainment parameter $E$ is given by
\begin{equation}
E = \frac{c_m}{\Pr_T} (\Ri_{\max} - \Ri).
\end{equation}

Using the empirically determined values $c_\varepsilon=0.21$, $c_m=0.25$ and $\Pr_T=0.8$, we find that
\begin{equation}
  E = 0.31 (\Ri_{\max} - \Ri), \quad \quad \Ri_{\max} = 0.15.
\end{equation}

Interestingly,
this entrainment prediction is in very good agreement with the
simulation results for the low angle cases as is clear from figure
\ref{fig:alpha_E}(b),
strongly suggesting that this model is appropriate, and that there
is indeed a  \rev{(particular kind of)} `switch-off' of entrainment in these flows for
sufficiently large $\Ri$.
Crucially, \rev{this particular} `switch-off' is not associated with a suppression of the
turbulence: for all cases, the turbulence levels $e_T$ were
similar, \rev{and finite.}  Rather it is the scaled \emph{intensity} of the turbulence
$e_T^{1/2}/u_T$ which tends to zero as $\Ri \rightarrow
\Ri_{\max}$. This key phenomenology is captured  by the theoretical
model, which predicts that for $\alpha \ll 1$, $e_T \sim B_0$ and
$u_T \sim B_0^{1/2} \alpha^{-1/2}$, from which it follows that
$e_T^{1/2} / u_T \rightarrow 0$ in the limit of $\alpha \rightarrow
0$. This in itself can be understood by the fact that the flow will
keep accelerating \rev{such that the characteristic integral velocity
  scale $u_T$ defined in 
(\ref{eq:integrals2}) increases} until the shear is sufficiently large to produce a
sufficient amount of turbulence to be steady.
\rev{It is apparent that} there is a critical
Richardson number beyond which this steady balance cannot be attained.
\rev{Once again, it is important to remember that this `switch off' is
  not associated with a complete suppression of turbulence, but rather
  that the turbulence becomes proportionally very small compared
  to the characteristic scale of velocity of the current. Therefore,
  though  analyses of such quantities are beyond the scope of this paper,  other quantities associated
  with turbulence, such as the buoyancy flux and turbulent drag also
  remain finite as this `switch off' is approached.
  Critically, however, the central point remains that quasi-steady turbulence cannot
  apparently
  be sustained for flows with characteristic Richardson number larger
  than the critical value of $\Ri_{\max}=0.15$.}

\section{Discussion}\label{sec:disc}

As already noted, there is a clear analogy of the temporal gravity current with the behaviour observed in
stratified plane Couette flow
by \cite{Deusebio2015} and \cite{Zhou2017},
where the mean flow is observed to adjust so that
$\Ri \lesssim 0.2$.  In that flow
geometry, there is by construction a
constant (with distance from the walls) vertical heat flux,
leading to self-similarity in the mean velocity and buoyancy  profiles
which can be then described well
using classical Monin-Obukhov similarity theory.
Importantly, the observation that there
is a maximum possible Richardson number in this flow
irrespective of the externally
imposed parameter values (e.g. of the flow's Reynolds number)
arises in the limit where the interior of the flow
is  assumed to be not \emph{directly} affected by the near-wall dynamics,
suggesting some possibility for generalisation, although it
is also important to appreciate that the
specific numerical value of this maximum possible
Richardson number
is determined using empirical constants
essential to the Monin-Obukhov theory. A key characteristic is that
the velocity and buoyancy mean profiles can be well-modelled as linear
functions of height, and that characteristic appears
also to be observed in the simulations reported here.

The key parameter to compare these temporal gravity current flows with 
stratified plane Couette flows, as
discussed in \S \ref{sec:intro}, is the Obukhov length, which
characterises
the relative importance of shear production and vertical buoyancy flux. 
For the flows under consideration here, a  
local Obukhov scale can be calculated, following \cite{Nieuwstadt1984}, as
\begin{equation}
L_o = -\frac{\overline{w'u'} | \overline{w'u'}
  |^{1/2}}{\overline{w'b'}}.
\label{eq:lodef}
\end{equation}
For scales smaller than $L_o$, the dynamics are expected
to be dominated by shear, while scales larger than $L_o$ are
expected to be increasingly affected by the (in this context
stabilising) effects of buoyancy.
The prediction from the model developed in \S \ref{sec:model} for this scale is
\begin{equation}
\begin{split}
L_o &= \frac{K_m}{K_\rho}\frac{\hat{S} \cos{\alpha}}{\hat{N}^2} K_m^{1/2} \hat{S}^{1/2} 
    = \frac{8}{9} Pr_T^2 \widehat{\Ri}_f^{-1} \cos{\alpha} \left(\frac{9}{8} a_e
      c_m \right)^{1/2} a h,
      \end{split}
\label{eq:Lodef}
\end{equation}
where $e$ has been replaced by its  maximum value (with $z$), which occurs at $z/h=\eta_I/2$, where $e = 9/8 a_e B_0^{1/2}$.
%

\begin{figure}
\centering
\includegraphics{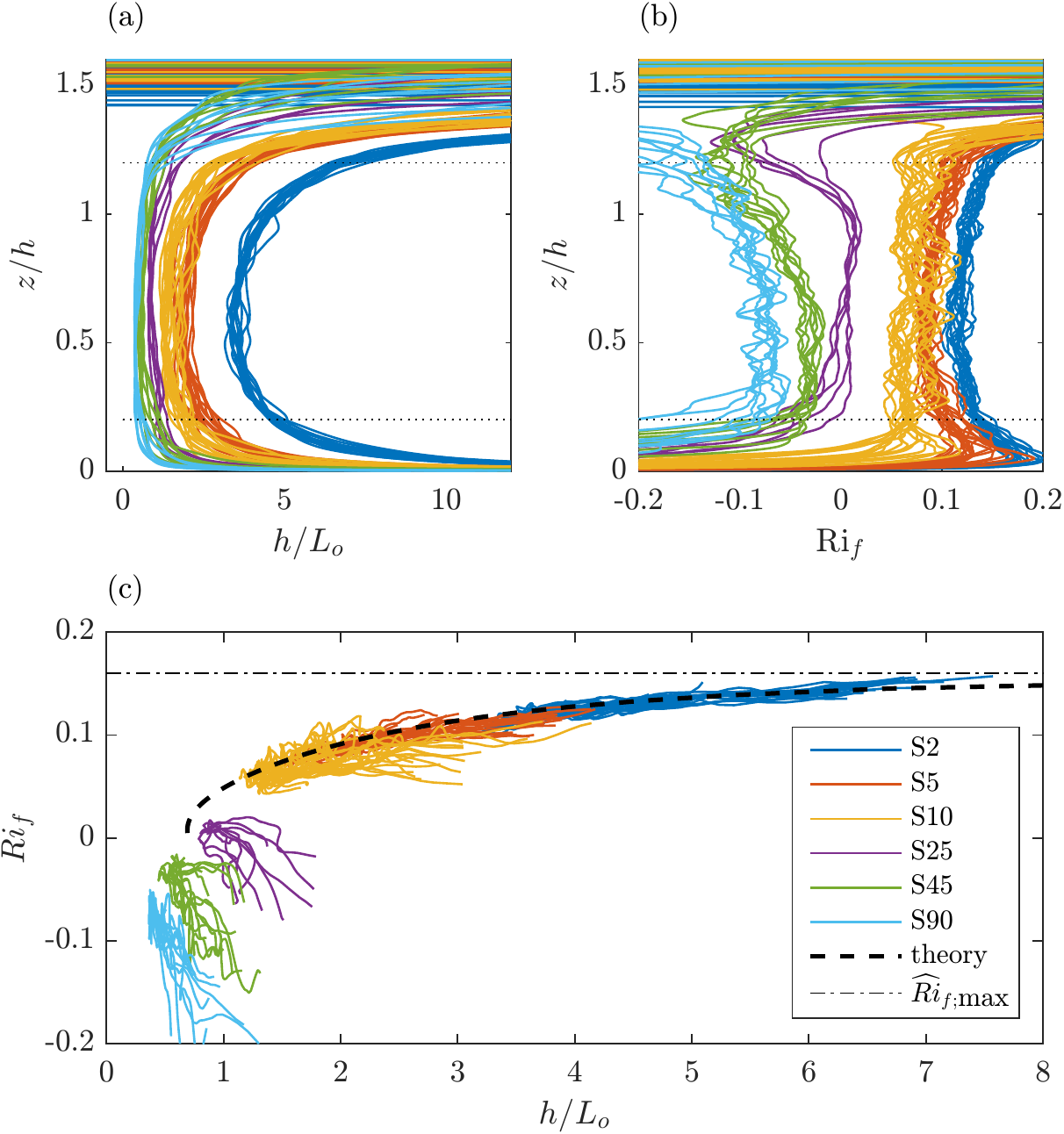}
\caption{a) Variation with $z/h$ of $h/L_o$, as defined in
  (\ref{eq:Lodef});
b) Variation with $z/h$ of $\Ri_f$ as defined in
(\ref{eq:rigrifgamdef});
c) Variation of $\Ri_f$ with $h/L_o$ for various simulations, as
denoted in the legend. The theoretical prediction (\ref{eq:Lodef}) is
plotted with a dashed line, and the theoretical maximum value
$\Ri_{f,max}$ is plotted with a dot-dashed line. \rev{For the time interval over which results are presented see Table \ref{tab:simdata} .} }
\label{fig:alpha_Lo}
\end{figure}

In figure \ref{fig:alpha_Lo}(a), the variation 
with $z/h$ of $h/L_o$ is plotted for the various
simulations. It is apparent that $h/L_o$ is close to constant across
the main part of the current, and also $h/L_o$ increases as $\alpha$
decreases, 
implying naturally that buoyancy becomes more significant across the main depth of the current as the slope angle decreases.
This observation is further confirmed by considering the variation with depth of $\Ri_f$ for the same simulations, as shown in figure \ref{fig:alpha_Lo}(b). 
Here, the near-constancy of $\Ri_f$ as a function of $z$ can be understood by the fact that $\overline{u}$ is approximately linear, and that $\overline{w'u'}$ and $\overline{w'b'}$ have a similar (approximately parabolic) shape, resulting in $\Ri_f \approx \textrm{constant}$ from small $\alpha$.
The increase of $\Ri_f$ to a maximum value $\alpha \rightarrow 0$ is consistent with figure \ref{fig:alpha_alphadep}.
In figure \ref{fig:alpha_Lo}(c), $\Ri_f(z)$ is plotted against $h/L_o(z)$ for $0.2 < z/h < 1.2$ (the interval is denoted by the dashed lines in figure \ref{fig:alpha_Lo}(a,b)) for all cases. 
For the flows with smaller angles $\alpha$, these two quantities prove to be highly correlated,
with excellent agreement to the theoretical prediction given by
(\ref{eq:Lodef}). \rev{From the definition of the flux Richardson
  number (\ref{eq:rigrifgamdef}), it is thus apparent that such flows
  continue to have non-zero buoyancy flux and turbulent drag as
  $\alpha \rightarrow 0$.}

The correlation displayed in figure \ref{fig:alpha_Lo}(c) is highly reminiscent of the behaviour of stratified plane Couette flow, as shown for example in figure 7
of \cite{Zhou2017}. 
In particular, both flows demonstrate the fact that 
as $h/L_o$ diverges, and so stratification becomes
increasingly important, $\Ri_f$ (and $\Ri$ from
(\ref{eq:rifrigrelate}) and the observation that $Pr_T \rightarrow 0.8$)) approach a finite value
$\Ri_{f,\max} \simeq 0.16$, largely consistent with the upper bound proposed, on semi-empirical grounds by \cite{osborn80}.

It is important to appreciate however that the connection is not
complete. In the \rev{temporal} gravity current flows considered here, we  do not observe a
transition
to intermittency as $\Ri$ increases above its critical value, in
contrast
\rev{to} the dynamics observed
in stratified plane Couette flow by \cite{Deusebio2015}.
Rather the \rev{temporal} gravity current flows still sustain turbulence, just the entrainment ceases to be
significant.
This might indicate a difference between flows driven via a body force or via boundary forcing, with temporal gravity currents falling in the former and stratified Couette flow falling in the latter category.

Another important issue to consider is that the observed numerical value  of $Ri_{max} \simeq 0.15$
is quite close to $1/4$,
the value  at the heart of the well-known Miles-Howard criterion for
linear normal mode stability of inviscid parallel steady stratified
shear flows, and consistent with  observations of $\Ri$ close to
1/4 in the equatorial undercurrent \citep{Smyth2013}.
\cite{Thorpe2009} conjectured that
this is not a coincidence, with the
flow being in a state of `marginal' stability. As soon as the mean
flow
properties are perturbed such that the Richardson number drops below
$1/4$,
shear instabilities are triggered which, through enhanced dissipation
and mixing, push the mean flow properties back towards conditions
corresponding to $\Ri \simeq 1/4$. 

Such a `kind of equilibrium' 
was actually also hypothesised by \cite{Turner1973}, and is 
also at least somewhat related to the concept of `self-organized
criticality' discussed in \S \ref{sec:intro}. At this stage, it is not possible
to confirm or disprove \rev{the conjecture of the relevance of the
  Miles-Howard criterion}, although
within these particular gravity current flows, there are at least two
characteristics which challenge
this conjecture. First,  there is always turbulence present, and so in
particular
the mean profiles are never actually realised precisely by the
flow, calling into question whether it is actually appropriate
to consider their linear stability. Second, we do not observe the
assumed
`bursting' of shear instabilities as the Richardson number drops below
some value instantaneously, rather the flow sustains turbulence
in a regime where $\Ri$ remains sufficiently small, and nontrivially
smaller than $1/4$ in point of fact. 

It also is interesting to note that \cite{Krug2015} reported an
equilibrium Richardson number of approximately 0.1 for two spatially
developing gravity currents with different initial Richardson number
$\Ri(0)$ and observed a cyclic buildup of velocity and density gradients
followed by their destruction by mixing and molecular diffusion. This
is somewhat reminiscent of  Thorpe's marginal stability
interpretation, even though, as is the case in the present study, the
flow never relaminarizes. The cyclic behavior observed in Krug et
al. (2015) exhibits local bursts of vigorous turbulence followed by
more quiescent periods. For the stronger stratification case, the flow
experiences stronger excursions (i.e. higher gradients), but flow
states inevitably revolved about the same central equilibrium of
buoyancy versus shear.

Nevertheless, despite these caveats,  
the simulations presented here constitute further evidence 
that such an approach to a relatively weakly stratified equilibrium
for  stratified shear flows may well be 
a generic property of turbulent stratified flows. 
This has profound implications for the parameterization and 
interpretation 
of mixing, and certainly suggests that the classical 
hypotheses of \cite{Ellison1957} concerning the accessibility of large values of the
turbulent 
Prandtl number should be revisited critically. 

\section{Conclusions}\label{sec:conc}

In this paper, we have revisited the classical entrainment experiments of \cite{Ellison1959} for gravity currents on inclined slopes,
focussing exclusively on the shear-driven entrainment and mixing
across
the `top' of the gravity current. 
 Using direct numerical simulations that are run for durations long
 enough for the flow to reach both  self-similarity in mean profiles  of velocity and buoyancy, and a dynamical equilibrium between the
competing effects of buoyancy and shear, 
we can demonstrate that the net effect of inner layer processes on entrainment and mixing in the outer layer is very small.
In all simulations we found that the turbulence is in a local
equilibrium in that \rev{turbulence production (primarily due to shear) is in equilibrium with dissipation and TKE}
transport is insignificant. 
We observed that for the simulations with $\alpha \le 10^o$ the turbulent diffusivity and dissipation rate can be parameterised using the turbulence kinetic energy $e$ and strain rate $S$, which can be used to predict the mixing parameters $\Ri_f$, $\Ri_g$ and $\Gamma$ as a function of $\alpha$. 


Both the simulations and the model point to a critical
Richardson number $\Ri_\textrm{max}$ as $\alpha \rightarrow 0$, which cannot be exceeded.
The model predicts an entrainment law $E=0.31 (\Ri_\textrm{max} - \Ri)$, where $\Ri_\textrm{max} \simeq 0.15$ which is in good agreement with our numerical results. 
This entrainment law hence predicts that $E\rightarrow 0$, i.e.\ entrainment switches off,  for $\Ri \rightarrow \Ri_\textrm{max}$.
One of the main objectives of the paper was to understand better what happens at this critical $\Ri_{\max}$ when $E \rightarrow 0$. 
We showed that under these conditions, although the entrainment
apparently `switches off', the  turbulence itself does not switch off,
but rather becomes
\rev{effectively} insignificant. 
Thus, even though $\Gamma$, $\Ri_f$ and
indeed even $\Ri_g$ become maximal as the slope angle $\alpha
\rightarrow 0$, the entrainment itself becomes higher order, and effectively negligible. 
This is because the \emph{mean} flow parameters, i.e. $S$ and $N$
become larger. This is due to the body force \rev{in our particular
  flow}, since as
\rev{$\alpha$ decreases to very small values} the flow continues to accelerate until the shear is sufficiently large to produce a sufficient amount of turbulence. 
Thus, quantities scaled with mean flow parameters like $e_T / u_T^2 \rightarrow 0$, and hence the turbulence becomes ineffective. 
Ellison's conjecture was that turbulence could continue to  be maintained at large $\Ri$ and thus,
since $\Ri_f \leq 1$, from (\ref{eq:rifrigrelate}), 
the turbulent Prandtl $Pr_T$ has to diverge in order to switch off
entrainment. 
We showed that turbulence indeed remains active as 
$\Ri \rightarrow \Ri_{\max}$, but it becomes insignificant and $Pr_T
\simeq 0.8$ remains finite.

\rev{In a fundamental fluid dynamical context, the findings in this paper are consistent with those of previous work which focussed on the small-space aspects associated with the outer interface separating turbulent and non-turbulent regions that controls
the mass flux of outer fluid into the turbulent region. 
This so-called turbulent-nonturbulent interface (TNTI) involves viscous diffusion of
vorticity at a rate that is governed by the local energy dissipation rate
and kinematic viscosity. The TNTI drives a local entrainment velocity
$v_n$ that is proportional to the Kolmogorov velocity scale $u_{\eta}$
over its convoluted surface area $A_\eta$. We can write the entrainment
rate \citep{Krug2013, vanReeuwijk2017} as
$E=v_nA_\eta/(u_TA)$, which involves the product of the ratio between the local
entrainment velocity and outer velocity scale and the ratio between the 
surface area of the TNTI and its projected area $A$. Both ratios were
shown to decrease with stratification level in \cite{Krug2013} and \cite{vanReeuwijk2017} hence depleting $E$. The present study
suggests that, at critical $Ri$, $v_n/u_T \rightarrow 0$ because turbulence intensity goes to zero, while we also have that $A_\eta/A\rightarrow 1$ because the scaling exponent of the TNTI goes to zero \citep{Krug2017}.
The latter is caused by the fact that the vertical dimension of the interface scales as $e^{1/2}/S$ for this flow, which tends to zero as $\Ri \rightarrow \Ri_\text{max}$.
Thus, the micro-scale and macro-scale viewpoints are fully consistent,
as one would expect them to be.}

We conjecture that at least some of the aspects of this flow
may be generic  among different stratified flows (including for
example plane Couette flow). The critical aspect appears to be that  the velocity field and buoyancy
field adjust so that they reach a balance and direct effects of
buoyancy production on turbulence and mixing are relatively small,
thus leading to mixing properties 
consistent with Osborn's classical mixing parameterisation, and a
turbulent Prandtl number $Pr_T \sim O(1)$. In particular, there is no
evidence
of scaled mixing, quantified by such quantities as the flux Richardson
number $\Ri_f$ and the 
turbulent flux coefficient $\Gamma$, exhibiting non-monotonic variation
with stratification as commonly observed
experimentally \citep{Linden1979}. Just as in plane Couette
flow \citep{Zhou2017}, 
such flows can only access the weakly stratified `left flank' of such
mixing curves.   \rev{In the nomenclature of \cite{Mater2014}, these
  flows
  are `shear-dominated', and indeed they appear never to be able
  to be in the `buoyancy-dominated' regime.} It is
still unclear why the particular value of $\Ri_\textrm{max} \simeq 0.15$ is
selected as the critical value, \rev{in particular as to whether it
  has any relationship to the Miles-Howard criterion,} and further research is undoubtedly
needed to understand why this value arises, and why the mixing
dynamics
appears to be inevitably located on this \rev{inherently
  `shear-dominated'} `left flank'.

\rev{Furthermore, it is also important to remember that in this paper we have
  focussed exclusively on one specific entrainment and mixing
  mechanism associated with gravity currents, the shear-driven mixing
  at the top of a current, driven by buoyancy when propagating down a 
  slope of finite, yet potentially small, angle.
 Naturally, in geophysical contexts, other processes are significant,
 and may indeed be leading order. Such processes include  of course,
 turbulence induced by other mechanisms, for example
 bottom roughness, tidal motions, and breaking waves. Furthermore,
 mixing may be dominated by the dynamics in the immediate vicinity of
 the gravity current's head, while the current itself may be forced by
 pressure
 gradients or indeed other mechanisms not directly related to the
 along-slope component of the buoyancy force, which slope is often
 extremely small, and generically not constant. The results presented
 here could thus be interpreted as demonstrating when the potential
 significance of such
 shear-driven entrainment should be considered when assembling a full
 picture of the entrainment and mixing dynamics of geophysically
 relevant
 gravity current flows.}

\section*{Acknowledgements}
The computations were made possible by an EPSRC Archer Leadership grant, the UK Turbulence Consortium under grant number EP/R029326/1 and the excellent HPC facilities available at Imperial College London.
The research activity of C.P.C. was supported by EPSRC Programme grant EP/K034529/1 entitled ‘Mathematical Underpinnings of  Stratified Turbulence’. 
This research was also supported in part by the National Science Foundation under grant no.\ NSF PHY17-48958. 
The hospitality of the Kavli Institute of Theoretical Physics (KITP) at the University of California, Santa Barbara, during the TRANSTURB17  programme is gratefully acknowledged.

\mbox{}

\nomenclature{$E$}{Entrainment coefficient [-] }
\nomenclature{$t$}{Time [s]}
\nomenclature{$x$}{Streamwise coordinate [m]}
\nomenclature{$y$}{Spanwise coordinate [m]}
\nomenclature{$z$}{Wall-normal coordinate [m]}
\nomenclature{$u$}{Streamwise velocity [ms$^{-1}$]}
\nomenclature{$v$}{Spanwise velocity [ms$^{-1}$]}
\nomenclature{$w$}{Wall-normal velocity [ms$^{-1}$]}
\nomenclature{$b$}{Buoyancy [ms$^{-2}$]}
\nomenclature{$\nu$}{Kinematic viscosity [m$^2$s$^{-1}$]}
\nomenclature{$h_0$}{Initial layer thickness [m]}
\nomenclature{$b_0$}{Initial buoyancy [ms$^{-2}$]}
\nomenclature{$U_0$}{Initial velocity [ms$^{-1}$]}
\nomenclature{$g$}{Gravitational acceleration [ms$^{-2}$]}
\nomenclature{$\rho$}{Fluid density [kgm$^{-3}$]}
\nomenclature{$\rho_a$}{Ambient fluid density [kgm$^{-3}$]}
\nomenclature{$\alpha$}{Slope angle [-]}
\nomenclature{$u_T$}{Characteristic velocity [ms$^{-1}$]}
\nomenclature{$b_T$}{Characteristic buoyancy [ms$^{-2}$]}
\nomenclature{$e_T$}{Characteristic turbulence kinetic energy [m$^2$s$^{-2}$]}
\nomenclature{$\lambda_T$}{Characteristic Taylor length scale [m]}
\nomenclature{$\varepsilon_T$}{Characteristic dissipation rate [m$^2$s$^{-3}$]}
\nomenclature{$h$}{Layer thickness [m]}
\nomenclature{$B_0$}{Buoyancy integral [m$^2$s$^{-2}$]}
\nomenclature{$Re$}{Reynolds number [-]}
\nomenclature{$Pr$}{Prandtl number [-]}
\nomenclature{$Ri$}{Bulk Richardson number [-]}
\nomenclature{$Re_0$}{Initial Reynolds number [-]}
\nomenclature{$Ri_0$}{Initial bulk Richardson number [-]}
\nomenclature{$Re_b$}{Buoyancy Reynolds number [-]}
\nomenclature{$Re_\lambda$}{Taylor Reynolds number [-]}
\nomenclature{$N$}{Buoyancy frequency [s$^{-1}$]}
\nomenclature{$t_{run}$}{Simulation time [s]}
\nomenclature{$t_*$}{Turnover time [s]}
\nomenclature{$t_{stat}$}{Time interval over which statistics are collected [s]}
\nomenclature{$P_S$}{TKE production by shear [m$^2$s$^{-3}$]}
\nomenclature{$P_B$}{TKE production by buoyancy [m$^2$s$^{-3}$]}
\nomenclature{$\varepsilon$}{Dissipation rate [m$^2$s$^{-3}$]}
\nomenclature{$Ri_f$}{Flux Richardson number [-]}
\nomenclature{$Ri_g$}{Gradient Richardson number [-]}
\nomenclature{$\Gamma$}{Turbulent flux coefficient [-]}
\nomenclature{$\mathcal{P}_S$}{Integral TKE production by shear [m$^3$s$^{-3}$]}
\nomenclature{$\mathcal{P}_B$}{Integral TKE production by buoyancy [m$^3$s$^{-3}$]}
\nomenclature{$\mathcal{E}$}{Integral dissipation rate [m$^3$s$^{-3}$]}
\nomenclature{$K_m$}{Eddy viscosity [m$^2$s$^{-1}$]}
\nomenclature{$K_\rho$}{Eddy diffusivity [m$^2$s$^{-1}$]}
\nomenclature{$Pr_T$}{Turbulent Prandtl number [-]}
\nomenclature{$a$}{Coefficient for layer thickness [-]}
\nomenclature{$a_u$}{Coefficient for characteristic streamwise velocity [-]}
\nomenclature{$a_b$}{Coefficient for characteristic buoyancy [-]}
\nomenclature{$Ri_{\max}$}{Maximum bulk Richardson number [-]}
\nomenclature{$Ri_{f;\max}$}{Maximum flux Richardson number [-]}
\nomenclature{$Ri_{g;\max}$}{Maximum gradient Richardson number [-]}
\nomenclature{$\Gamma_{\max}$}{Maximum turbulent flux coefficient [-]}
\nomenclature{$c_m$}{Coefficient for eddy viscosity [-]}
\nomenclature{$c_\rho$}{Coefficient for eddy diffusivity [-]}\nomenclature{$c_\varepsilon$}{Coefficient for dissipation rate [-]}\nomenclature{$S$}{Strain rate [s$^{-1}$]}
\nomenclature{$\widehat{Ri}_f$}{Outer layer flux Richardson number [-]}
\nomenclature{$\widehat{Ri}_g$}{Outer layer gradient Richardson number [-]}
\nomenclature{$\widehat{Ri}$}{Outer layer bulk Richardson number [-]}
\nomenclature{$\widehat{\Gamma}$}{Outer layer turbulent flux coefficient [-]}
\nomenclature{$\eta$}{Similarity variable [-]}
\nomenclature{$\eta_1$}{Edge of outer layer [-]}
\nomenclature{$f_u$}{Similarity profile for $u$ [-]}
\nomenclature{$f_b$}{Similarity profile for $b$ [-]}
\nomenclature{$f_e$}{Similarity profile for $e$ [-]}
\nomenclature{$\hat{N}$}{Outer layer buoyancy frequency [s$^{-1}$]}
\nomenclature{$\hat{S}$}{Outer layer strain rate [s$^{-1}$]}
\nomenclature{$L_o$}{Obukhov length [m]}
\printnomenclature

\bibliographystyle{jfm}

\end{document}